\documentclass[twocolumn,showpacs,preprintnumbers,amsmath,amssymb]{revtex4}
\usepackage{graphicx}% Include figure files
\usepackage{dcolumn}% Align table columns on decimal point
\usepackage{bm}% bold math
\usepackage{float}

\usepackage{color}
\usepackage{refstyle}
\usepackage{mathrsfs}
\usepackage{esint}
\usepackage[unicode=true,pdfusetitle,bookmarks=true,
bookmarksnumbered=false,bookmarksopen=false,breaklinks=false,
pdfborder={0 0 0},backref=false,colorlinks=true,citecolor=red]
{hyperref}
\usepackage{cleveref}
\providecommand\eqref[1]{\ref{eq:#1}}
\renewcommand\b[1]{{\bf  #1}}
\renewcommand\epsilon{\varepsilon}

\renewcommand\phi{\varphi}

\newcommand\del{\nabla}
\newcommand\dd{\mathrm{d}}
\renewcommand\v{\mathsf{v}}
\renewcommand\u{\mathsf{u}}

\DeclareMathOperator{\sgn}{sgn}
\allowdisplaybreaks[3]

\begin{document}
\title{Topological Sound and Flocking on Curved Surfaces}
%\author{Suraj Shankar\\[5pt]
%	Department of Physics\\
%\emph{Syracuse University}\\[20pt]}
\author{Suraj Shankar$^{a,b}$}
\author{Mark J. Bowick$^{a,b}$}
\author{M. Cristina Marchetti$^{a,b}$}
%\email{}
\affiliation{$^a$Physics Department and Syracuse Soft Matter Program, Syracuse University, Syracuse, NY 13244, USA.\\
$^b$Kavli Institute for Theoretical Physics, University of California, Santa Barbara, CA 93106, USA.}
%\author{M. Cristina Marchetti}
%\affiliation{Department of Physics, Syracuse University}
%\email[]{sushanka@syr.edu}
\date{\today}
%\clearpage
\begin{abstract}
	Active systems on curved geometries are ubiquitous in the living world. In the presence of curvature orientationally ordered polar flocks are forced to be inhomogeneous, often requiring  the presence of topological defects even in the steady state due to the constraints imposed by the topology of the underlying surface. In the presence of spontaneous flow the system additionally supports long-wavelength propagating sound modes which get gapped by the curvature of the underlying substrate. We analytically compute the steady state profile of an active polar flock on a two-sphere and a catenoid, and show that curvature and active flow together result in symmetry protected topological modes that get localized to special geodesics on the surface (the equator or the neck respectively). These modes are the analogue of edge states in electronic quantum Hall systems and provide unidirectional channels for information transport in the flock, robust against disorder and backscattering.
\end{abstract}
\maketitle
%\pagebreak
%\begin{multicols}{2}
\section{Introduction}
Flocking, defined as the self-organized ordered motion of collections of self-propelled units \cite{vicsek1995novel,toner2005hydrodynamics}, has been the flagship of active matter for some time now \cite{ramaswamy2010mechanics,marchetti2013hydrodynamics}. Active entities dissipate energy to perform work and generate motion, leading to {\em sustained} and {\em local} breaking of detailed balance. Examples abound in the living world, ranging from bird flocks \cite{cavagna2010scale} to bacterial suspensions \cite{sokolov2007concentration} and migrating cells \cite{szabo2006phase}, and include synthetic analogues, such as reconstituted cytoskeletal extracts \cite{schaller2010polar,sanchez2012spontaneous,keber2014topology}, vibrated granular media \cite{deseigne2010collective,narayan2007long} and self-propelled colloids \cite{palacci2013living}. Active matter as a field aims to provide broad organizational principles applicable to a wide class of these non-equilibrium systems over many scales.

Collective cellular motion on curved surfaces is ubiquitous in developmental processes, such as morphogenesis and embryonic development~\cite{streichan2017quantification,vasiev2010modeling,ewald2008collective}, or when cells migrate in the gut \cite{fatehullah2013cell,ritsma2014intestinal} or on the surface of the growing cornea \cite{collinson2002clonal}, and also affects cell division \cite{mishra2012cylindrical}.
Recent \textit{in vitro} work has demonstrated a direct effect of substrate curvature on cytoskeletal alignment and cell motility in epithelial cells \cite{yevick2015architecture}.
Understanding the behavior of  active matter on curved surfaces or confined by curved boundaries. is therefore timely.  There has been growing recent interest in understanding this at a fundamental level, with the focus divided between the effect of curved confining walls on so-called \emph{scalar} (non-aligning) active matter \cite{fily2014dynamics,fily2015dynamics,fily2016equilibrium,nikola2016active,smallenburg2015swim} and on aligning active matter systems, of either \emph{nematic} \cite{green2016geometry,keber2014topology,sanchez2012spontaneous} or \emph{polar} \cite{fily2016active,sknepnek2015active} symmetry. Even at the level of non-interacting self-propelled particles, the curvature of confining walls can yield  surprising features, such as inhomogeneous density and pressure profiles \cite{nikola2016active,fily2014dynamics} and the breakdown of an equilbrium interpretation \cite{fily2016equilibrium,fily2015dynamics}. In the presence of aligning interactions that promote orientational order, curvature has an even more profound effect since it frustrates such order, often requiring topological defects \cite{nelson2002defects} that in active systems become dynamical and are capable of driving spatio-temporal patterns and complex motion \cite{keber2014topology}. With flexible walls or membranes present, activity can lead to spontaneous motion and rectification \cite{maitra2014activating,mallory2014curvature}.

A generic property of the ordered state of polar active matter is spontaneous flow  and thus the breaking of time-reversal symmetry \cite{marchetti2013hydrodynamics}. It is known that carefully engineered lattice structures with flows induced either spontaneously by activity~\cite{souslov2016topological}, or through an external drive~\cite{yang2015topological}, can host exotic unidirectional sound modes that are localized at the edges of the sample and topologically protected. The presence of topologically protected edge states  in classical phononic \cite{prodan2009topological,kane2014topological} and photonic \cite{raghu2008analogs} systems has lead to extensive exploration of topological meta-materials, with properties akin to electronic topological insulators and quantum Hall states \cite{hasan2010colloquium}. Here we show that a polar active fluid on a curved substrate supports similar topologically protected modes, even in the absence of any underlying periodicity or lattice structure. This should be contrasted with many of the systems considered previously which required a carefully-designed meta-material structured on an artificial lattice. The phenomenon reported here is akin to the one recently found in geophysical flows of oceans or the earth's atmosphere, where equatorially trapped Kelvin waves were reinterpreted as topologically protected modes~\cite{marston2017}. In that case the flow is imposed externally by the earth's rotation. In our active fluid, in contrast, flow occurs with no external drive, resulting in \emph{spontaneous} topologically protected modes.

The presence of these topological modes relies on three important ingredients:
\begin{itemize}
	\item the spontaneous polar order and associated flow that breaks time reversal symmetry;
	\item the fact that in polar active fluids the order parameter also plays the role of a flow velocity, resulting in distinctly non-equilibrium self-advection not present in equilibrium polar fluids~\cite{marchetti2013hydrodynamics}; 
	\item the non-zero gaussian curvature of the underlying substrate.
\end{itemize}
We emphasize that the long-wavelength topologically protected modes discussed here are \emph{generic}, in the sense that they occur for active polar flow on any curved surface of non-vanishing gaussian curvature. Recent work has considered active polar patterns on a cylinder \cite{srivastava2013patterning}. In this case the gaussian curvature vanishes and there are consequently no topologically protected sound modes.
In the following we demonstrate the phenomenon explicitly for flocking on the sphere, which has constant positive curvature, and on the catenoid, which has negative, spatially varying curvature.

In Sec.~\ref{sec:sphere}, we analyze the continuum Toner-Tu model of flocking on a sphere and analytically compute the steady state configuration of the polar ordered flock. Due to the curvature, the ordered state is forced to be inhomogeneous in general. On a sphere polar order additionally requires topological defects or vortices (the ``hairy-ball theorem'' \cite{eisenberg1979proof}) in order to satisfy the constraints imposed by the global topology of the surface. With a well motivated ansatz, we show that the covariant hydrodynamic model is capable of predicting generic inhomogeneous steady ordered phases that accommodate the curvature and topology of the underlying surface in a natural fashion. The steady flocking state on the sphere corresponds to the rotating band seen in recent particle simulations by~\citet{sknepnek2015active} and is a novel find in itself as it is peculiar to the active system (a passive polar liquid crystal on a sphere would have a very different equilibrium profile).

Having obtained a steady ordered state, in Sec.~\ref{subsec:sound} we examine its excitations. Even though our system is overdamped due to the presence of a substrate, the ordered polar flock supports long-wavelength propagating sound modes \cite{tu1998sound}. The presence of curvature introduces an additional length scale in the problem and gaps the sound spectrum at long wavelengths. In other words the propagation rate of long-wavelength density fluctuations is finite. This is a distinct property of  active polar fluids and it arises because the polarization field plays the dual role of the order parameter and flow velocity and is therefore subject to the same lensing effect that forces flow to move along geodesics on curved surfaces~\cite{sknepnek2015active,marston2017}. With a spectral gap opened, in Sec.~\ref{sec:top} we show that a polar ordered flock on a curved surface supports topologically protected sound modes that are localized to special geodesic curves on the surface (at which the gap in the spectrum closes). In Sec.~\ref{sec:neg}, we compute the steady state of an ordered polar flock on a catenoid and show that topological modes are also present on this negative-curvature surface. In contrast to the setup of \citet{souslov2016topological}, where the lattice structure was instrumental (along with the active flow) in generating a gapped spectrum at intermediate frequencies, here the curvature does the job though now at long wavelengths, implying that the result is quite general.

%\vspace{2em}
\section{Toner-Tu equations on a curved surface}

We consider an active polar fluid on a $2d$ surface. To make generic predictions independent of specific microscopic realizations, we work in the continuum limit and use the well-tested hydrodynamic description of a fluid of overdamped self-propelled particles provided by the Toner-Tu equations \cite{toner1995long,toner1998flocks,toner2005hydrodynamics}, appropriately modified to account for the curvature of the sphere \cite{fily2016active}. Mass conservation implies a continuity equation for the density field, $\rho$,
\begin{equation}
	\partial_t\rho+\del_{\mu}p^{\mu}=0\;,\label{eq:cont}
\end{equation}
with $\mu=\theta,\phi$ and $\b{p}=\rho\b{u}$ the polarization density of the active fluid. Due to the presence of a substrate, momentum is not conserved and the particles’ velocity is assumed to be aligned with their direction of self-propulsion, leading to the identification of $\b{u}$ with the flow velocity of the active fluid. Note that on a curved surface parallel transport of vectors requires the use of covariant derivatives \cite{do1992riemannian},
\begin{equation}
	\del_{\mu}p^{\nu}=\partial_{\mu}p^{\nu}+\Gamma^{\nu}_{\alpha\mu}p^{\alpha}\;,
\end{equation}
where $\Gamma^{\nu}_{\alpha\mu}$ are the appropriate Christoffel symbols. The equation for the polarization density is given by
\begin{widetext}
\vspace{-1em}
	\begin{equation}
	\partial_tp^{\mu}+\lambda p^{\nu}\del_{\nu}p^{\mu}=\left[a(\rho-\rho_c)-b\ g_{\alpha\beta}p^{\alpha}p^{\beta}\right]p^{\mu}+\nu\left(\Delta p^{\mu}+K_G p^{\mu}\right)+\nu'\del^{\mu}\del_{\nu}p^{\nu}-v_1\del^{\mu}\rho\;.\label{eq:peq}
\end{equation}
\end{widetext}
Note that $\b{u}$ here plays the dual role of an order parameter field (polarization) and velocity, as discussed in \cite{marchetti2013hydrodynamics}.
The transport coefficients $\nu$ and $\nu'$ are the shear and bulk viscosities (or anisotropic elastic constants), respectively, $a,b>0$ are coefficients setting the magnitude of the mean field polarized state for $\rho>\rho_c$ (the critical density for the flocking transition), $\lambda$ is a kinematic convective parameter and $v_1>0$ is a compressional modulus. The last term on the right hand side of Eq.~(\ref{eq:peq}) is the first term in a density expansion of the gradient of the swim pressure \cite{yang2014aggregation,takatori2014swim} that describes the flux of propulsive forces across a unit plane of material. There are other nonlinear terms in the original Toner-Tu equations, but we only retain the most important ones here. In particular, we keep the convective nonlinearity $\lambda\b{p}\cdot\del\b{p}$ that is responsible for long-ranged order in $2d$ \cite{toner1995long,toner2012reanalysis} and the leading density dependence in the symmetry breaking ($a\rho\b{p}$) and pressure like terms ($v_1\del\rho$) that lead to dynamical self-regulation \cite{gopinath2012dynamical}, phase-separation \cite{bertin2009hydrodynamic,solon2015pattern,caussin2014emergent} and long-wavelength instabilities of the ordered phase \cite{mishra2010fluctuations,bertin2006boltzmann}. The absence of Galilean invariance means that $\lambda\neq 1/\rho$. Additional nonlinear advective terms $\sim \lambda_2\b{p}\del\cdot\b{p}$, $\lambda_3\del|\b{p}|^2$ are also present in general, but do not qualitatively change our results below (see Appendix~(\ref{sec:lambda}) for an analysis with $\lambda_2,\lambda_3\neq 0$).

Curvature enters Eq.~(\ref{eq:peq}) in two crucial places (apart from the covariant derivatives): (i) the cubic term setting the magnitude of the polarization explicitly involves the metric tensor $\mathfrak{g}$ ($|\b{p}|^2=g_{\alpha\beta}p^{\alpha}p^{\beta}$) and (ii) the gaussian curvature $K_G$ explicitly appears in the viscous term because the strain rate tensor is a symmetrized derivative of the velocity and the covariant derivatives do not commute. The presence of $K_G\neq 0$ is a direct dynamical consequence of the Poincar{\'e}-Hopf theorem \cite{eisenberg1979proof} from which it follows that topological defects or vortices are required to accommodate vector order on a curved closed surface like the sphere. A covariant hydrodynamic treatment of active fluids on a curved surface has also been developed by \citet{fily2016active}. These authors derived the continuum equations by coarse-graining a microscopic model of self-propelled particles, which allowed an explicit computation of the transport coefficients in terms of microscopic parameters. The form of the continuum equations obtained in Ref.~\cite{fily2016active} is identical to those used here, the only distinction being that $\mathcal{O}(\del^2)$ terms are neglected in that work, including the explicit $K_G$ term. In the following, we shall similarly neglect $\del^2$ terms.
%one cannot have perfect homogeneous vector order, leading inevitably to the formation of topological defects or vortices.

\section{Polar flock on a sphere}
\label{sec:sphere}
\begin{figure}[]
	\centering
	\includegraphics[width=0.35\textwidth]{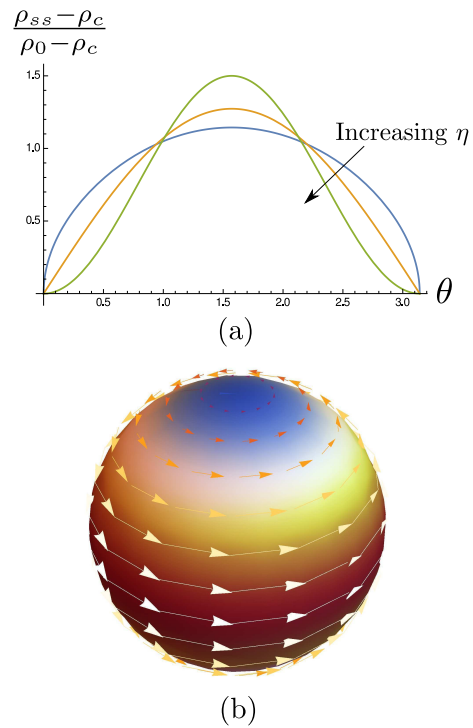}
	\caption{(a) The normalized density profile of a polar flock on a sphere given in Eq.~(\ref{eq:rhoss}), for $\eta=0.5$ (blue), $1$ (orange) and $2$ (green). (b) The density and polarization profiles for $\eta=2$, now shown on the sphere. The color describes the density from the maximum (red) at the center of the polar band to $\rho_c$ (blue) at the poles of the sphere. The polarization also vanishes at the poles.}
	\label{fig:density}
\end{figure}

As an example of a curved surface with constant positive curvature, we consider an active polar flock on the surface of a sphere of radius $R$. In local spherical polar coordinates $\{\theta,\phi\}$, the canonical metric and curvature on $S^2$ are
\begin{equation}
	\mathfrak{g}=R^2(\dd\theta\otimes\dd\theta+\sin^2\theta\dd\phi\otimes\dd\phi)\;,\quad\quad K_G=\dfrac{1}{R^2}\;.
\end{equation}
The only non-vanishing Christoffel symbols are
\begin{equation}
	\Gamma^{\theta}_{\phi\phi}=-\sin\theta\cos\theta\quad\mathrm{and}\quad\Gamma^{\phi}_{\theta\phi}=\cot\theta\;.
\end{equation}

\subsection{The steady state of a polar flock on the sphere}
At low mean density ($\rho_0<\rho_c$), the isotropic phase with constant density and $\b{p}=\b{0}$ is stable. For $\rho_0>\rho_c$, where the mean-field solution in flat space is a state of constant density  and finite, but uniform polarization, on the sphere one obtains polar, spatially varying states. 
%Going across the mean field transition, in the ordered phase, the steady state polarization is no longer uniform and spatially homogeneous. 
Since the particle number is conserved, there can be no sinks or sources of flow. The simplest configuration allowed by the required conservation of topological defect charge that must sum up to the Euler characteristic $\chi=2$ of the sphere is then a circulating band wrapping around an equator, with two vortices of charge $+1$ at the two opposing poles. This yields a density band with polarization in the azimuthal direction and both density and polarization vanishing at the poles, consistent with the band state  reported recently in simulations of  polar particles on the sphere~\cite{sknepnek2015active}.
% (i.e. along the $\partial_{\phi}$ vector field, which is a Killing field on the round sphere) 
%leading to an azimuthally polarized state, with the polarization and the density vanishing at the poles. 

An explicit solution can be found analytically by assuming azimuthal symmetry, with $\rho=\rho_{ss}(\theta)$, $p^{\theta}=0$ and $p^{\phi}=p_{ss}^{\phi}(\theta)$.
The continuity equation is then satisfied identically. To simplify the polarization equation, we neglect the viscous terms as they are higher order in gradients (suppressed by $1/R^2$) compared to the other terms arising from self-propulsion. In the microscopic realization of self-propelled polar particles with repulsive short range forces and aligning interaction studied in Ref.~\cite{sknepnek2015active} this approximation corresponds to the regime where inter-particle repulsion (contributing to pressure) and active self-propulsion dominate over viscous and elastic stresses. We then neglect the laplacian terms entirely by setting $\nu=0$ (the bulk viscosity $\nu'$ drops out with our assumption of azimuthal symmetry). This leaves us with
\begin{gather}
	\lambda\sin\theta\cos\theta\left(p_{ss}^{\phi}\right)^2=\dfrac{v_1}{R^2}\partial_{\theta}\rho_{ss}\;,\label{eq:ptheta}\\
	p_{ss}^{\phi}[a(\rho_{ss}-\rho_c)-bR^2\sin^2\theta\left(p_{ss}^{\phi}\right)^2]=0\;.\label{eq:pphi}
\end{gather}
Writing $X(\theta)=\rho_{ss}(\theta)-\rho_c$, and seeking a solution with $p_{ss}^{\phi}\neq0$ we can eliminate $p_{ss}^{\phi}$ from the two equations to obtain
\begin{equation}
	\dfrac{\dd X}{\dd\theta}=\left(\dfrac{a\lambda}{bv_1}\right)\cot\theta X\implies X(\theta)=X(\pi/2)(\sin\theta)^{\eta}\;,
\end{equation}
where
\begin{equation}
\eta=\dfrac{\lambda a}{b v_1}
\end{equation}
is a dimensionless parameter that controls the shape of the solution, with $\eta>0$ for the density profile to be a physical solution 
%(see Appendix~\ref{sec:lambda} for the modification to $\eta$ when $\lambda_2,\lambda_3\neq 0$). 
(note that $\sin\theta>0$ over the entire range $\theta\in[0,\pi]$). By symmetry the density will be maximum at $\theta=\pi/2$ (the equator). Letting $\rho_{ss}(\pi/2)=\rho_{\mathrm{max}}$ we find
\begin{equation}
	\rho_{ss}(\theta)=\rho_c+\left(\rho_{\mathrm{max}}-\rho_c\right)\sin^{\eta}\theta\;.
	\label{eq:rhoss1}
\end{equation}
We stress that the dependence on $R$ has dropped out from Eq.~(\ref{eq:rhoss1}), which therefore represents a universal density profile for an ordered flock on any size sphere. Finally, we express  $\rho_{\mathrm{max}}$ in terms of the average density $\rho_0\equiv\langle\rho_{ss}\rangle$ by requiring
\begin{gather}
	\rho_0=\dfrac{R^2}{4\pi R^2}\int_{0}^{2\pi}\dd\phi\int_0^{\pi}\dd\theta\sin\theta\rho_{ss}(\theta)\;.
\end{gather}
%with the result
%\begin{gather}	
%    \rho_0=\rho_c+\left(\rho_{\mathrm{max}}-\rho_c\right)\dfrac{\sqrt{\pi}}{2}\dfrac{\Gamma\left(1+\dfrac{\eta}{2}\right)}{\Gamma\left(\dfrac{3+\eta}{2}\right)}.
%\end{gather}
%Inverting this in terms of $\rho_0$, 
to obtain the final expression for the density profile as
\begin{equation}
	\rho_{ss}(\theta)=\rho_c+(\rho_0-\rho_c)A_{\eta}\sin^{\eta}\theta\;,\label{eq:rhoss}
\end{equation}
with $A_{\eta}=2\Gamma((3+\eta)/2)/[\sqrt{\pi}\Gamma(1+\eta/2)]$.
%\dfrac{2}{\sqrt{\pi}}\dfrac{\Gamma\left(\dfrac{3+\eta}{2}\right)}{\Gamma\left(1+\dfrac{\eta}{2}\right)}$.
In order for this density profile to exist, we additionally require that $|{\bf p}_{ss}|^2>0$ and obtain
\begin{gather}
%	|{\bf p}_{ss}|^2=R^2\sin^2\theta(p_{ss}^{\phi})^2=\dfrac{a}{b}(\rho_{ss}(\theta)-\rho_c),\\
|{\bf p}_{ss}|^2=\dfrac{a}{b}(\rho_0-\rho_c)A_{\eta}\sin^{\eta}\theta\;.
\end{gather}
As expected, an ordered flock only exists for $\rho_0>\rho_c$, and the magnitude  
of the steady state polarization and the density have the same inhomogeneous profile as shown in Fig.~\ref{fig:density}a,
%as the density and  below which we have the homogenous disordered phase with $\rho_{ss}=\rho_0$ and ${\bf p}_{ss}={\bf 0}$. 
with the direction of polarization chosen spontaneously.

This band solution is unrelated to the traveling bands found in flat space \cite{caussin2014emergent,solon2015pattern,mishra2010fluctuations}, which occur close to the mean-field transition and are absent deep in the ordered phase. The inhomogeneous solution obtained is simply the ordered flocking state on a sphere. The spatially inhomogeneous profile arises from the interplay of mass fluxes ($\sim v_1\del\rho$) and convective fluxes ($\sim\lambda\b{p}\cdot\del\b{p}$) that cannot be driven to zero on a curved surface. Hence the spatial inhomogeneity is made inevitable by curvature.
\begin{figure}[]
	\centering
	\includegraphics[width=0.49\textwidth]{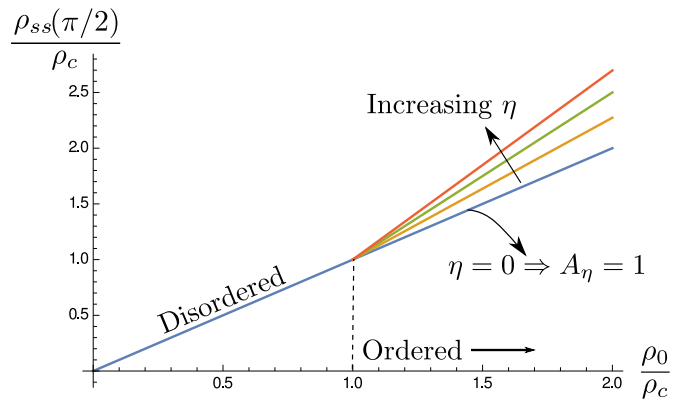}
	\caption{The peak (maximum) density at the equator on a sphere, as we vary the mean density $\rho_0$. For $\rho_0<\rho_c$, we are in the disordered phase with $\rho_{ss}(\pi/2)=\rho_0$. For $\rho_0>\rho_c$, we have a polar band with the density profile given in Fig.~\ref{fig:density}. The density reaches its maximum at the center of the band and grows with $\rho_0$, with a slope $A_{\eta}>1$. When the convective parameter $\lambda\to0$, $\eta\to0$, resulting in $A_{\eta}\to1$ and we go back to the homogeneous profile as in the flat plane.}
	\label{fig:eqrho}
\end{figure}
The present solution is expected to break down within a region of angular width $\theta_{\mathrm{m}}\sim\exp(-a\rho_cR^2/\nu)$ (hence exponentially small on a large sphere)
around the poles of the sphere, at the core of the vortices, as the elastic stresses will become important at short scales. 
%The maximum angular width of the equatorial band around the poles within which the effect of viscous or elastic stresses are important goes like $\theta_{\mathrm{m}}\sim\exp(-a\rho_cR^2/\nu)$ and is exponentially small on a large sphere. 
%Hence, staying away from the poles on a large enough sphere, the density profile obtained in Eq.~(\ref{eq:rhoss}) is a sufficiently good approximation, with the leading finite size correction only being a shift in the maximum density at the equator ($\rho_{\mathrm{max}}\to\rho_{\mathrm{max}}+\nu\eta/2aR^2$). 
Therefore the profile obtained is a robust and universal prediction of the continuum theory, similar to the rotating band seen in particle simulations of Ref.~\cite{sknepnek2015active}.
%, although the dependence of the exponent $\eta$ on the microscopic parameters has not been verified.

For an equilibrium polar or ferroelectric liquid crystal (say, a compressible lyotropic smectic-C film \cite{de1997physics}), where the polarization is strictly an order parameter field and does not play the role of a velocity, the important convective nonlinearity in Eq.~(\ref{eq:peq}) is absent ($\lambda=0$, though $\lambda_2$ and $\lambda_3$ can be present \cite{kung2006hydrodynamics}) and the band solution we have here is absent ($\eta=0$). In this case, even in the ordered phase, the density remains homogeneous and on a large enough sphere, we have nearly uniform polar order everywhere ($|\b{p}_{ss}|\simeq$ const.), but for two isolated defects at the poles, whose core size  $\xi\sim\sqrt{\nu/a(\rho_0-\rho_c)}$ is a microscopic length scale deep into the ordered state.

\subsection{Linearizing about the steady state}
\label{subsec:sound}
Well below the mean-field transition ($\rho_0<\rho_c$), the isotropic disordered state ($\rho_{ss}=\rho_0$ and $\b{p}_{ss}=\b{0}$) is linearly stable at long wavelengths with fluctuations in the polarization relaxing quickly and density perturbations relaxing diffusively at long time, just as in the plane \cite{baskaran2008hydrodynamics}. The curvature does not affect the disordered phase. It is only in the ordered state that we find novel excitations with non-trivial topological properties.

Here we consider the linear dynamics of small amplitude perturbations about the steady ordered state flock, letting $\rho=\rho_{ss}(\theta)+\delta\rho$ and $\b{p}=\b{p}_{ss}(\theta)+\delta\b{p}$. We focus on the long-wavelength propagating sound modes that are present even in the plane for an ordered flock \cite{tu1998sound}, and continue to neglect all the viscous and elastic coupling. These are higher order in gradients and only give rise to damping of the sound modes. As the base state we are linearizing about is inhomogeneous, we additionally confine ourselves to a tangent plane linearization about a fixed latitude away from the poles (a preferred local coordinate system is picked out spontaneously by the polar order allowing for an unambiguous notion of latitude). Setting $\theta=\theta_0+y$ for a given latitude $\theta_0$, with $\theta_0<\pi/2$ corresponding to the northern hemisphere and $\theta_0>\pi/2$ to the southern hemisphere, relabelling $\phi$ as $x$, and letting $\delta p^{\theta}\to\v$ and $\delta p^{\phi}\to\u$, we obtain
\begin{align}
	&\partial_t\delta\rho+\partial_x\u+\partial_y\v+\v\cot\theta_0=0\;,\label{eq:linsphere1}\\
	&\partial_t\u+\lambda p_0\partial_x\u+\dfrac{v_1}{R^2\sin^2\theta_0}\partial_x\delta\rho\nonumber\\
				&\quad=p_0(a\delta\rho-2bR^2p_0\sin^2\theta_0\ \u)-\v\dfrac{\lambda p_0(\eta+2)}{2}\cot\theta_0\;,\label{eq:linsphere2}\\
			 &\partial_t\v+\lambda p_0\partial_x\v+\dfrac{v_1}{R^2}\partial_y\delta\rho=2\u\lambda p_0\sin^2\theta_0\cot\theta_0\;,\label{eq:linsphere3}
\end{align}
with $p_0=p^{\phi}_{ss}(\theta_0)$ the azimuthal polarization at latitude $\theta_0$, which is finite as long as we are away from the poles, $\theta_0\neq0,\pi$.
% and correspondingly $p_0\neq 0$ too. 
Note that stability of the steady state requires $\eta>0$, hence $\lambda>0$. 
%So all the parameters appearing in the above equations are positive. 
The only terms that can be negative are those proportional to $\cot\theta_0$, arising from the Christoffel symbols, which changes sign as one crosses the equator at $\theta_0=\pi/2$. 
\begin{figure*}[]
	\centering
	\includegraphics[width=\textwidth]{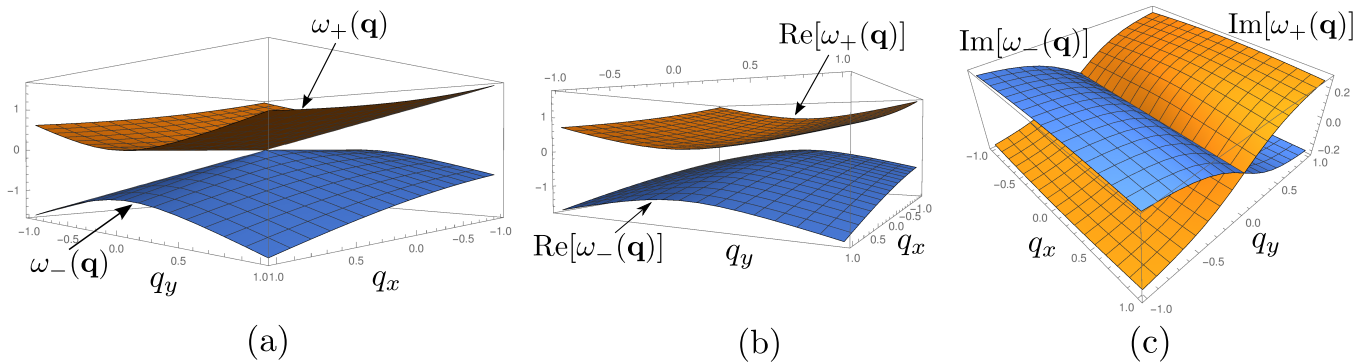}
	\caption{The relevant (slow) sound modes. (a) $m=0$ and the gap between the two bands is closed. (b) and (c) are the dispersion bands for $m=0.2$ (Eq.~(\ref{eq:wpm})) and we directly see that a gap has opened in the real part of the spectrum, while the imaginary part of the frequency has a single crossing line at $q_y=0$. The variables are chosen to be $\theta_0\simeq78^{\circ}$, $\alpha\simeq2.03$, $\bar{\lambda}\simeq1.03$ and $\beta=\bar{v}_1=1$.}
	\label{fig:freq}
\end{figure*}

Next we perform a Galilean boost to a comoving frame by letting $x\to x-\lambda p_0 t$ (comoving with the logitudinal sound and \emph{not} the flock itself), and relabel $\bar{\lambda}=\lambda p_0$, $\alpha=a p_0>0$, $\beta=2b p_0^2 R^2\sin^2\theta_0>0$, $\bar{v}_1=v_1/R^2>0$ and
\begin{equation}
m=-\cot\theta_0\;.
\end{equation}
The redefined parameters are summarized in Table~\ref{table:params}. Since the flock breaks Galilean invariance, this is not a symmetry operation, and yields
\begin{gather}
	\partial_t\delta\rho-\bar{\lambda}\partial_x\delta\rho+\partial_x\u+\partial_y\v=m\ \v\;,\\
	\partial_t\u+\dfrac{\bar{v}_1}{\sin^2\theta_0}\partial_x\delta\rho=\alpha\delta\rho-\beta\u+\dfrac{\bar{\lambda}(\eta+2)}{2}m\ \v\;,\\
	\partial_t\v+\bar{v}_1\partial_y\delta\rho=-2\bar{\lambda}\sin^2\theta_0 m\ \u\;.
\end{gather}
Here $m$ is a constant of fixed sign at any given non-equatorial latitude and changes sign across the equator, with $m<0$ in the northern hemisphere and $m>0$ in the southern half, vanishing only at the equator where $\theta_0=\pi/2$. We show below that a non-vanishing value of $m$ leads to a band gap (Fig.~\ref{fig:freq}b and~\ref{fig:freq}c) in the sound mode spectrum that acquires the necessary structure for non-trivial band topology. This, along with the vanishing of $m$ at the equator, naturally suggests that the equator behaves as a ``boundary'' between two different ``bulk'' media (the northern and southern hemispheres), thereby allowing for localized topologically protected excitations on it.
\begin{table}[H]
	\centering
	{
	\setlength{\extrarowheight}{8pt}
	\begin{tabular}{|c|c|c|c|}
		\hline
		$\lambda$ & $a$ & $b$ & $v_1$\\
		\hline
		$~\bar{\lambda}=\lambda p_0~$ & $~\alpha=a p_0~$ & $~\beta=2b p_0^2 R^2\sin^2\theta_0~$ & $~\bar{v}_1=v_1/R^2~$\\ [1ex]
		\hline
	\end{tabular}
	}
	\caption{A summary of the parameters redefinitions in the model.}
	\label{table:params}
\end{table}
To simplify the notation we let $|\delta\Psi\rangle\equiv(\delta\rho,\u,\v)$ and recast the linearized equations that control the linear stability of the steady state in the form of a Schr{\"o}dinger like equation (in Fourier space, with $\Psi(\b{q})=\int\dd^2r\ e^{-i\b{q}\cdot\b{r}}\Psi(\b{r})$), as
\begin{gather}
	i\partial_t|\delta\Psi\rangle=H|\delta\Psi\rangle\;,\\
	\quad H(\b{q})=\left(\begin{matrix}
			-\bar{\lambda} q_x & q_x & i m+q_y\\
			i\alpha+\dfrac{\bar{v}_1q_x}{\sin^2\theta_0} & -i\beta & im\bar{\lambda}\left(\dfrac{\eta}{2}+1\right)\\
			\bar{v}_1q_y & -2im\bar{\lambda}\sin^2\theta_0 & 0
	\end{matrix}\right)\;.
\end{gather}
The eigenvalues of $H(\b{q})$ directly give the sound mode frequencies ($|\delta\Psi\rangle\propto e^{-i\omega t}$). An important distinction compared to the Schr{\"o}dinger equation is that the matrix $H$ is not hermitian and therefore the linearized mode spectrum is not purely real, due to dissipative terms describing the overdamped dynamics and the absence of Galilean invariance. For $m=0$ ($\theta_0=\pi/2$) the equations reduce to those of the planar case. Fluctuations in the polarization magnitude ($\u$) are controlled by a fast mode that decays on microscopic time scales $\sim\beta^{-1}$,
\begin{equation}
	i\omega_0(\b{q})=\beta-i\dfrac{\alpha}{\beta}q_x+\mathcal{O}(q^2)\;.
\end{equation}
The density ($\delta\rho$) and the transverse goldstone mode ($\v$) are the only slow modes that remain propagating at long wavelengths (as $\b{q}\to 0$),
\begin{equation}
	\omega_{\pm}(\b{q})=\dfrac{1}{2\beta}\left[(\alpha-\beta\bar{\lambda})q_x\pm\sqrt{(\alpha-\beta\bar{\lambda})^2q_x^2+4\bar{v}_1\beta^2q_y^2}\right]\;,
\end{equation}
where we have only kept terms to leading order in $\b{q}$.

For non-zero but small $m\neq 0$ ($\theta_0\neq\pi/2$), corresponding to the regions close to the equator in either hemisphere, the dispersion relations  can be written as
\begin{widetext}
	\vspace{-1em}
\begin{gather}
	i\omega_0(\b{q})=\beta-i\dfrac{\alpha}{\beta}(q_x+\dfrac{2m\bar{\lambda}}{\beta}\sin^2\theta_0q_y)+\mathcal{O}(q^2)\;,\\
	\omega_{\pm}(\b{q})=\dfrac{1}{2\beta}\left[q_x(\alpha-\beta\bar{\lambda})\pm\sqrt{q_x^2(\alpha-\beta\bar{\lambda})^2+4\beta(m-i q_y)(2m\alpha\bar{\lambda}\sin^2\theta_0+i\bar{v}_1\beta q_y)}\right]+\mathcal{O}(m^2,q^2,mq)\;.\label{eq:wpm}
\end{gather}
\end{widetext}
In the next section we analyze this mode structure.

\section{Topological sound}
\label{sec:top}
We immediately see that the two branches of the propagating modes given by Eq.~(\ref{eq:wpm}) have a gap at $\b{q}=0$ of width $\Delta=|\omega_+(0)-\omega_{-}(0)|$ proportional to $|m|$, with
\begin{equation}
	\Delta=2|m|\sin\theta_0\sqrt{\dfrac{2\alpha\bar{\lambda}}{\beta}}+\mathcal{O}(m^3)\;.
\end{equation}
The terms explicitly involving $m$ in the dispersion relations, responsible for the opening of the gap, are obtained only in the presence of \emph{both} curvature \emph{and} spontaneous active flow. In the plane \emph{static} long-wavelength deformation of both the density and the broken symmetry mode leave the system unchanged. On a curved surface, in contrast, spatially uniform deformations of either ``slow'' field ($\delta\rho$ and $\v$) cannot be static and invariably lead to dynamics in the system. As a result of curvature-induced forces, long-wavelength deformations of would be slow modes are required to have a finite frequency, resulting generically in the $\b{q}=0$ gap of the sound spectrum, in sharp contrast to the conventional behavior of hydrodynamics in flat geometry \cite{martin1972unified}.

It is useful to compare the effect at hand with one that occurs in geophysical flows. In a frame comoving with the flock, the finite curvature of the sphere plays a role similar to the Coriolis force that would be present for a passive fluid on a rotating sphere. In the case of the earth's atmosphere, this has recently been shown to give a gapped sound spectrum and equatorially confined Kelvin and Yanai waves \cite{gill2016atmosphere} that are topological in origin \cite{marston2017}. In our active system no external flow or rotation needs to be imposed and the absence of Galilean invariance allows for independent tuning of the material parameters (such as $\lambda$) in order to probe regimes that are not accessible to passive fluids.

On times scales $t\gg\beta^{-1}$ we can slave the fast mode $\u$ to the slow fields, $\u\simeq\alpha\delta\rho/\beta+\mathcal{O}(\del\delta\rho)$ \footnote{The term involving $\bar{\lambda}(\eta+2)m\v$ is not relevant as its contribution to the dispersion is subdominant, as can be seen from its absence in Eq.~(\ref{eq:wpm}).}. Upon eliminating $\u$ we get a reduced set of dynamical equations involving only $\delta\rho$ and $\v$. After rescaling the wave-vector $|\alpha\beta-\bar{\lambda}|q_x/\beta\to q_x$, the linear matrix controlling the dynamics of $\delta\rho, \v$ fluctuations is given by 
\begin{equation}
	D(\b{q})=\left(\begin{matrix}
		s q_x & q_y+i m\\
			\bar{v}_1q_y-i\mu m & 0
	\end{matrix}\right)\;,\label{eq:D}
\end{equation}
where $\mu=2\bar{\lambda}\alpha\sin^2\theta_0/\beta>0$ ($\theta_0\neq0,\pi$) and $s=\sgn(\alpha-\bar{\lambda}\beta)$ ($s=0$, if $\bar{\lambda}=\alpha/\beta$). One can easily check that the eigen-frequencies of $D(\b{q})$ are exactly given by $\omega_{\pm}(\b{q})$ (Eq.~(\ref{eq:wpm})) modulo the appropriate rescaling of $q_x$. As $D(\b{q})$ is still non-hermitian, we need to evaluate right and left (adjoint) eigenvectors
\begin{align}
	D(\b{q})|\psi_i\rangle&=\omega_i(\b{q})|\psi_i\rangle\;,\\
	D^{\dagger}(\b{q})|\chi_i\rangle&=\omega^{\ast}_i(\b{q})|\chi_i\rangle\;,\quad (i=\pm)
\end{align}
with the biorthogonality relation $\langle\chi_i|\psi_j\rangle=\delta_{ij}$. It is important to keep in mind the regime in which $D(\b{q})$ provides a valid approximation to the complete dynamics. For
\begin{equation}
	\dfrac{1}{\beta}\ll t\ll\dfrac{\beta}{\bar{\lambda}^2\eta m^2}\;,\quad\dfrac{\bar{\lambda}|m|}{\beta}\ll 1\;,\label{eq:regime}
\end{equation}
we can neglect the fast $\u$ mode and not worry about higher order terms in both $\b{q}$ and $m$. This can be achieved deep in the ordered phase on a large enough sphere, in which case $\beta$ is large, allowing for a large window of time in which the dynamics is dominated by $D(\b{q})$. With this set of simplifications, the linear dynamical matrix is always diagonalizable and $\omega_+\neq\omega_-$ as long as $m\neq0$ or $\b{q}\neq0$ allowing one to adiabatically deform our model to have purely real eigenvalues by smoothly taking $\mu\to \bar{v}_1$  (for $\mu\neq 0$). In the process, the spectral gap remains open as long as $m\neq 0$.

\begin{figure*}[]
	\centering
	\includegraphics[width=\textwidth]{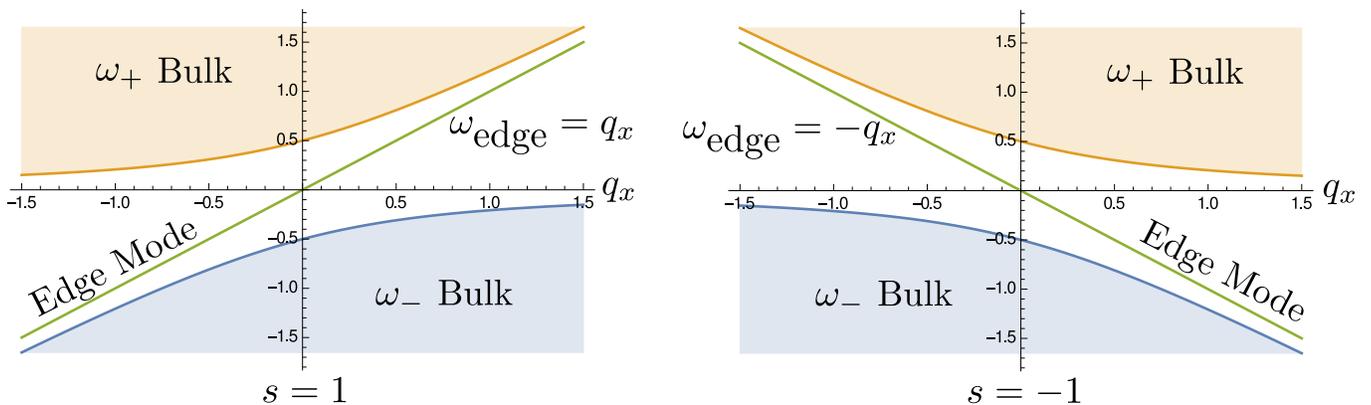}
	\caption{The bulk and edge mode spectrum for the case when $s=\pm1$ (shown for the simple case when $\mu=\bar{v}_1$) and $m(y)$ varying from $-1$ to $+1$.}
	\label{fig:edge}
\end{figure*}

In order to establish the topological nature of the band structure, we compute the associated $\mathsf{U}(1)$ Berry gauge connection and curvature \cite{berry1984quantal}
\begin{equation}
	\mathcal{A}_{\pm}=i\langle\chi_{\pm}|\del_{\b{q}}|\psi_{\pm}\rangle\ ,\quad \mathcal{F}_{\pm}(\b{q})=\del_{\b{q}}\times\mathcal{A}_{\pm}\ ,
\end{equation}
along with the Chern numbers \cite{nakahara2003geometry} for each band
\begin{equation}
	C_{\pm}=\int\dfrac{\dd^2q}{2\pi}\mathcal{F}_{\pm}=\pm\dfrac{s}{2}\sgn(m)\ ,\label{eq:chern}
\end{equation}
where $s=\sgn(\alpha-\bar{\lambda}\beta)$. The Chern number here is only quantized to a half integer as we work directly in the continuum long-wavelength approximation and the closing of the gap (for $m=0$) only gives rise to one Dirac cone like structure (this is the contribution to the ``parity anomaly'' or Hall conductance associated to a single Dirac cone in a Chern insulator \cite{haldane1988model}). An appropriate regularization for large $\b{q}$, guarantees the Chern number to be an integer \cite{hasan2010colloquium,ryu2010topological}. This calculation though still does have worth in predicting the correct number of topologically protected edge modes present when we stitch two of these regions with different Chern numbers together, via the bulk-edge correspondence. So each gap closing (change in sign of $m$), leads to a single localized edge mode, which as we shall see is unidirectional.

As anticipated earlier we find that the Chern number of the acoustic band is different in the northern ($m<0$) and southern ($m>0$) hemispheres, vanishing at the equator ($m=0$). Hence going across the equator, we have one gap closing (at $\b{q}=0$) with a band inversion, leading to a single topological sound excitation localized at the equator. Note that the Chern number also vanishes for $s=0$ which is obtained either when $p_0=0$ or $\bar{\lambda}=\alpha/\beta$, leading to a topologically trivial band structure. The first case corresponds to the absence of spontaneous active flow and the second to a partial restoration of Galilean invariance (in this limit, both density and goldstone mode excitations propagate with the same longitudinal speed). So the vanishing of the Chern number and associated band triviality for $s=0$ is \emph{not} due to the closure of a gap, but instead due to a restoration of symmetry. As $\beta=2bR^2p_0^2\sin^2\theta_0$ depends on the latitude at which we are, the vanishing of $s$ (for $\bar{\lambda}=\alpha/\beta$) is a condition on the polar angle $\theta_0$. There is a critical density
\begin{equation}
	\rho^{\ast}=\rho_c+\dfrac{1}{2A_{\eta}\lambda},
\end{equation}
such that for $\rho_c<\rho_0<\rho^{\ast}$, $s\neq0$ on the entire sphere. Deeper into the ordered state ($\rho_0>\rho^{\ast}$), there are two latitudes at angles $\theta_{\pm}$ such that $\sin^{\eta}\theta_{\pm}=(\rho^{\ast}-\rho_c)/(\rho_0-\rho_c)$, at which $s=0$. Even though the band topology changes as we cross the latitudes at $\theta_{\pm}$ (Eq.~(\ref{eq:chern})), the spectrum remains gapped throughout and hence we \emph{do not} have any gapless excitations localized at $\theta_{\pm}$. The change in the Chern number across these special latitudes is due to an accidental additional symmetry (Galilean invariance) instead of the gap closing, thereby circumventing the bulk-boundary correspondence. This is a well known point in quantum topological insulators, only realized here in a peculiar fashion as the protecting ``symmetry'' varies spatially in a single sample.

To summarize there are three crucial ingredients in this system that lead to and protect the topologically non-trivial band structure -
\begin{itemize}
	\item Breaking of time-reversal symmetry by the active polar flow. Changing the direction of spontaneous polarization (flow) changes the sign of $s$ (as $p_0\to-p_0$ leads to $\alpha\to-\alpha$, $\bar{\lambda}\to-\bar{\lambda}$, $\beta\to\beta$ and $\mu\to\mu$).
	\item The presence of the convective nonlinearity $\lambda\neq 0$; an equilibrium passive polar liquid crystal will not therefore exhibit these modes.
	\item The curvature of the base surface which opens up a gap in the sound spectrum. Changing the gaussian curvature exchanges the regions with positive and negative ``$m$''.
\end{itemize}

This is entirely analogous to the Haldane model \cite{haldane1988model} where the closing of the gap at the Dirac point is protected by time-reversal symmetry, which when broken by the local magnetic field leads to a band structure with a non-trivial topology. Though the active system is not hermitian with purely real frequencies, the structure of the localized equatorial mode for varying $m(y)$ is adiabatically connected to its hermitian analogue \cite{leykam2017edge}, the Jackiw-Rebbi soliton \cite{jackiw1976solitons}.

\begin{equation}
|\delta\Psi_{\mathrm{edge}}\rangle=\psi_0 e^{-\eta\int^y_0m(y')\dd y'+iq_x(x-st)}\left(\begin{matrix} 1\\ 0\end{matrix}\right),
\label{eq:psi}
\end{equation}
where $\psi_0$ is a normalization constant. The edge mode spectrum $\omega_{\mathrm{edge}}(q_x)=sq_x$ corresponds to a pure one-way density wave that connects the two bulk bands (see Fig.~(\ref{fig:edge})). This edge mode is valid when $m(y)\to\pm m_0$ ($m_0>0$) for $y\to\pm\infty$.

On the sphere, reverting back to angular coordinates $\{\theta,\phi\}$, $m(\theta)=-\cot\theta$, which is positive in the southern hemisphere for $\theta>\pi/2$ ($y>0$). This gives a chiral equatorial density mode ($\v=0$), which in the lab frame looks like
\begin{equation}
	\mathsf{\delta\rho}_{\mathrm{edge}}(\theta,\phi;t)=\sin^{\eta}\theta\sum_{n\geq 0}[\mathsf{a}_n\ e^{i n(\phi-\alpha t/\beta)}+\mathrm{c.c}],\label{eq:eqmode}
\end{equation}
where $\mathsf{a}_n$ are complex constants depending on the initial perturbation applied. A snapshot of this density mode is shown in Fig.~\ref{fig:sphcatmode} for $n=6$ (and all other $\mathsf{a}_n$ vanishing). Equation~(\ref{eq:eqmode}) defines a localization length $\ell_{\mathrm{loc}}=R/\eta$ set by the curvature and material parameters of the active fluid and essentially given by the ratio of longitudinal to transverse sound speeds. Note that the result given in Eqs.~(\ref{eq:psi}) and~(\ref{eq:eqmode})  applies for $\lambda_3=0$.
%$\mu/\bar{v}_1=\eta=\lambda a/bv_1$ (the same exponent as that of the steady state profile) and is 
% Including $\lambda_2,\lambda_3\neq0$ makes $\mu/\bar{v}_1\neq\eta$ (
The general case of $\lambda_2,\lambda_3\neq0$ is given in Appendix~\ref{sec:lambda} and yields a different localization width for the equatorial mode. This topological edge mode propagates unidirectionally in the direction of the flock and is robust to disorder because there are no reverse channels into which it can scatter (though  it will eventually dissipate due to viscous and elastic damping).
 Unlike a Galilean invariant fluid for which $\eta=1$,  here $\eta$ and therefore the localization length can be tuned  by varying the system's parameters, although  the shape of the steady state profile remains unchanged.
 
\section{Polar Flock on a Negative Curvature Surface}
\label{sec:neg}
The presence of such topological excitations is generic in that they will always be present when one has a polar flock on a curved surface that looks locally like a surface of revolution \footnote{More precisely, one needs a Killing field on the surface dictating a symmetry direction. One can always have the Christoffel symbols vanish at a given point in Riemann normal coordinates, but in the presence of a Killing field, this extends to an integral curve along which the Christoffel symbols vanish, changing sign as you cross it. This provides the necessary structure for the existence of such topological modes.}. We illustrate this point on a catenoid, a surface with non-constant negative gaussian curvature.

\begin{figure}[]
	\centering
	\includegraphics[width=0.35\textwidth]{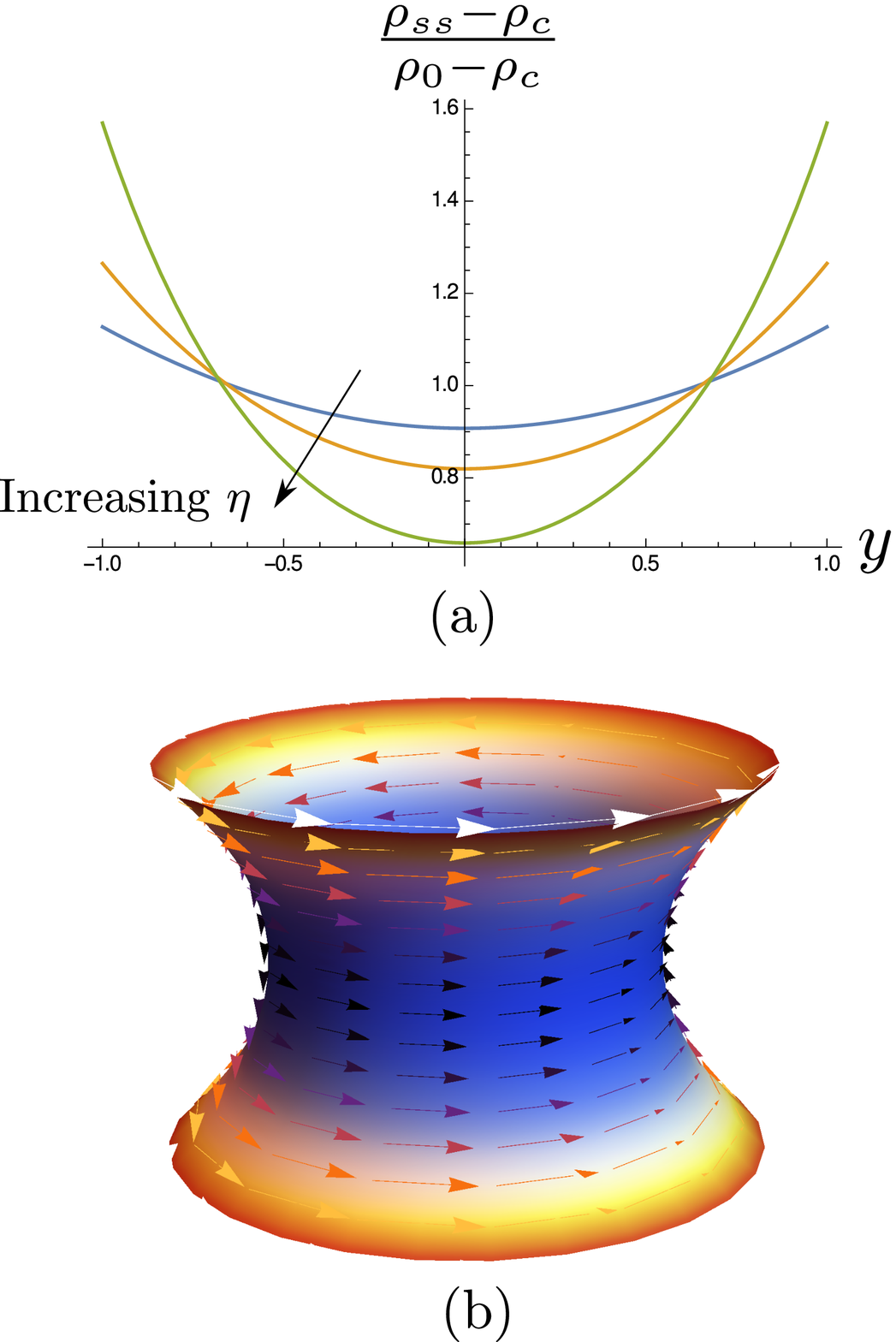}
	\caption{(a) The normalized density profile of a polar flock on a catenoid given in Eq.~\ref{eq:catrho}, for $\eta=0.5$ (blue), $1$ (orange) and $2$ (green). Note that, unlike the sphere, the density grows near the edge of the catenoid. (b) The density and polarization (for $\eta=2$) now shown on the catenoid. As before, blue corresponds to low density regions (at the neck) and red to high density.}
	\label{fig:catenoid}
\end{figure}

In local coordinates $\{y,\phi\}$ ($\phi$ once again being the periodic azimuthal direction), the metric and gaussian curvature on a catenoid are
\begin{gather}
	\mathfrak{g}=R^2\cosh^2y(\dd y\otimes\dd y+\dd\phi\otimes\dd\phi)\;,\\
	K_G(y)=-\dfrac{1}{R^2}\mathrm{sech}^4y\;,
\end{gather}
where $R$ is the radius of curvature at the neck of the catenoid. In contrast to the sphere, the gaussian curvature here is both negative and spatially varying. The only non-vanishing Christoffel symbols are
\begin{equation}
	\Gamma^{\phi}_{\phi y}=\Gamma^{y}_{yy}=-\Gamma^y_{\phi\phi}=\tanh y\;.
\end{equation}
Taking the same approach as for the sphere, neglecting viscous and elastic stresses, we consider an azimuthally symmetric ansatz for the steady state polar flock : $\rho=\rho_{ss}(y)$, $p^{y}=0$ and $p^{\phi}=p^{\phi}_{ss}(y)$. One can easily verify that  for $\rho_0>\rho_c$ the steady state density profile is then
\begin{equation}
	\rho_{ss}(y)=\rho_c+(\rho_0-\rho_c)B_{\eta}\cosh^{\eta} y\;,\label{eq:catrho}
\end{equation}
where $B_{\eta}<1$ is a constant that depends on $\eta$ and the height of the catenoid (which unlike the sphere is not compact and has to be taken finite). The details of the computation are given in Appendix.~\ref{sec:catenoid}.
In contrast to the sphere, which had a polar band with maximum density at the equator, the polar flock density is lowest at the neck of the catenoid ($y=0$), increasing on either side as one moves away from it.
The corresponding polarization profile is given by
\begin{equation}
	|\b{p}_{ss}|^2=\dfrac{a(\rho_0-\rho_c)}{b}B_{\eta}\cosh^{\eta}y\;.
\end{equation}
The density and polarization profiles are plotted in Fig.~\ref{fig:catenoid}a. Below the mean field transition ($\rho_0<\rho_c$), we recover the isotropic disordered phase ($\rho_{ss}=\rho_0$, $\b{p}_{ss}=\b{0}$).

Linearizing about this steady state, one finds that the equations governing the propagation of sound modes on the catenoid are essentially identical to that on the sphere (Eqs.~(\ref{eq:linsphere1}), (\ref{eq:linsphere2}), and (\ref{eq:linsphere3})), but with modified parameters. As a consequence of the negative curvature, the most important change is that $m=-2\tanh y$ is positive below the neck of the catenoid ($y<0$) and negative above ($y>0$), vanishing right at the neck ($y=0$).
This leads to a chiral mode of goldstone fluctuations localized at the neck of the catenoid ($\delta\rho=0$), which written in the lab frame is given by
\begin{equation}
	\delta\v_{\mathrm{edge}}(\phi,y;t)=\mathrm{sech}^{2}y\sum_{n\geq 0}\left[\mathsf{b}_n\ e^{in(\phi-\bar{\lambda}t)}+\mathrm{c.c}\right]\;,\label{eq:vedge}
\end{equation}
\begin{figure}[]
	\centering
	\includegraphics[width=0.5\textwidth]{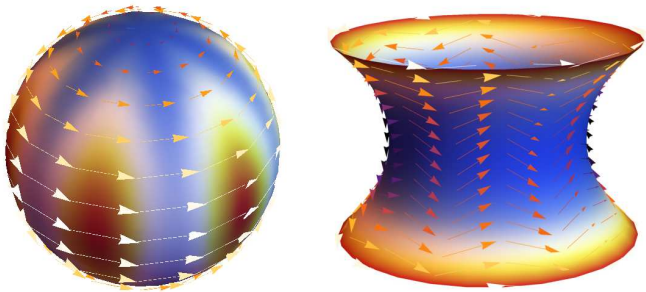}
	\caption{A representative snapshot of the equatorial density mode on a sphere (Eq.~(\ref{eq:eqmode})) and the localized goldstone mode on the catenoid (Eq.~(\ref{eq:vedge})). For clear visualization we have chosen the perturbation $\mathsf{a}_6,\mathsf{b}_6=0.5$ and all other $\mathsf{a}_n,\mathsf{b}_n=0$ ($n\neq 6$). We have also taken both $\eta=\mu/\bar{v}_1=2$ in both cases.}
	\label{fig:sphcatmode}
\end{figure}

where as before $\bar{\lambda}=\lambda p_0$  and $\mathsf{b}_n$ is a complex coefficient determined by the initial perturbation. A snapshot of this mode is shown as well in Fig.~\ref{fig:sphcatmode}. This mode propagates in the same direction as the flock, but with a different speed and is topologically protected. The localization length  $\ell_{\mathrm{loc}}=2R$ is controlled by the scale of the curvature in the system and seemingly independent of the material parameters of the flock.

\section{Conclusion}
The frustration associated with the interplay of curvature and order \cite{bowick2009two,turner2010vortices,nelson2002defects} has many consequences for crystals \cite{bowick2000interacting,bausch2003grain,irvine2010pleats,bowick2011crystalline,bendito2013crystalline}, tethered membranes \cite{nelson1987fluctuations,seung1988defects}, liquid crystalline membranes \cite{xing2012morphology,ramakrishnan2012role,mesarec2017curvature}, and jammed and glassy systems \cite{nelson1983liquids,modes2007hard}. The presence of activity adds an entirely new non-equilibrium dimension to the whole story, allowing for completely new physics arising from competing order, curvature and the active drive. The active polar fluid is peculiar as the polarization order parameter is also a velocity that advects the fluid around \cite{marchetti2013hydrodynamics}, with this dual role being at the heart of many of the phenomena we have explored in this article. In particular, unlike a superfluid film on a curved substrate (where the order parameter can be parallel transported trivially) \cite{vitelli2004anomalous}, the fact that the order parameter of the flock is a physical velocity implies that it advects itself non-trivially in the presence of curvature. This is nothing but a restatement of the physical fact that self-propelled particles move persistently along geodesics (in the absence of interactions), which get ``lensed'' by curvature, whereas passive polar particles don't do any such thing.

In order to handle these points we generalized the continuum Toner-Tu model for an active polar fluid, to an arbitrary curved surface and found new terms that are absent in flat space. In general, studying ordering phenomena on curved surfaces is rather complicated, even in equilibrium. This is when symmetry is often a very useful guiding tool, using which we explicitly computed the ordered phase of an active polar flock on two surfaces, the sphere and the catenoid. The continuum model affords us the privilege of having model independent predictions and we find many of the features of the steady ordered state of a polar flock are quite generic, with positive curvature surfaces having density profiles with a maximum, decreasing on either side of the peak, while negative curvature surfaces have the density profiles of the opposite kind, being minimal in the interior and increasing towards the boundary. Finding such spatially \emph{in}homogeneous exact solutions to the covariant Toner-Tu model is definitely a crucial starting point to being able to understand how the phenomenology of active matter in flat space translates to its curved variants.

In addition to the steady state with spontaneous flow, flocks in flat space also have dissipative sound modes with a linear (but angle dependent) dispersion \cite{tu1998sound}. It is here that all three players : curvature, order and activity come together with dramatic consequences. The presence of curvature gaps the sound mode spectrum at long wavelengths leading to a band structure with non-trivial topology protected by the broken time-reversal and Galilean symmetry in the system. We demonstrate this by computing the Chern number for the bands and show that this is a generic feature of flocks on curved surfaces. The most interesting result the non-trivial band topology has is to localize ``edge'' modes of density or goldstone mode fluctuations along special geodesics on the surface, at which the gap in the sound spectrum vanishes. The rather novel feature here is that the system is not artificially engineered as a metamaterial with some underlying lattice structure \cite{souslov2016topological,nash2015topological}, nor does it require any external forcing or fields of any kind \cite{yang2015topological}. The spontaneous flow is generated by activity breaking both time reversal and Galilean invariance simultaneously, while the curvature is responsible for the spectral gap in the ordered phase.

Topological excitations of the type described here are ``protected'' in the sense that they are robust against static perturbations and heterogeneities in the medium through which they propagate.  While quantifying the limits of such topological protection in active systems will require numerical work and remains to be explored, our work demonstrating that such topologically protected propagating modes are a \emph{generic} consequence of active flows on curved surfaces raises the question of whether nature may use this mechanism to guide and direct the robust transmission of intercellular physical forces in curved environments. 
It is therefore tempting to offer some speculation to the possible relevance of our findings to biology.
In a number of developmental phenomena, from wound healing to morphogenesis and organ development, living cells migrate collectively, offering an intriguing realization of a polar active fluid.  While a full understanding of the mechanisms that regulate collective cell migration is still out of reach, it is now widely recognized that the transmission of physical forces plays an important role, alongside biochemical signalling. For instance, propagating mechanical waves have been shown to mediate cooperative force transmission among epithelial cells in wound healing assays \cite{serra2012mechanical}. In many biological processes cell motion takes place on curved surfaces, as in cell renewal and repair in the highly folded intestine \cite{Holmberge201607260} and the shaping of the early limb bud in developing embryos \cite{hopyan2011budding}.
The effect of curvature on the dynamics of epithelial cells is  beginning to be explored \emph{in vitro} by examining collective cell migration on cylindrical capillaries of varying radii \cite{yevick2015architecture}. While cylinders have zero gaussian curvature which would not yield topologically protected states, these experiments clearly demonstrate that curvature affects cell morphology and dynamics by enhancing cell speed and cell extrusion. 
A more direct application of the work described here would be to a polar version of the active nematic vesicles described in Ref.~\cite{keber2014topology}. Here active vesicles were engineered by confining an active suspension of microtubule-kinesin bundles to the surface of a lipid vesicle. The interplay of activity and curvature yields a number of dynamical structures, including spontaneously oscillating defect textures and folding nematic bands, and ultimately activity-driven shape deformations of the vesicle. Our work may also be relevant to the physics of cell membranes that are activated through coupling to the polymerizing acto-myosin cortex, as modeled in recent work by \citet{maitra2014activating}.
 Finally, in the spirit of colloidal crystals on curved interfaces and reconstituted active systems, one might also envision synthetic experimental realizations of the topological sound modes investigated here, by depositing active Janus colloidal rods or active bio-filament motor complexes on the surface of a droplet or a vesicle shell.
 
\section{Acknowledgments}
We thank Brad Marston for inspiring this work through a beautiful talk delivered at KITP based on Ref.~\cite{marston2017}. We also thank Rastko Sknepnek and Vladimir Juri$\check{\mathrm{c}}$i\'{c} for useful discussions. This work was supported by the National Science Foundation at Syracuse University through awards DMR-1609208 (MCM \& SS) and DGE-1068780 (MCM) and at KITP under Grant No.~NSF PHY-1125915.
All authors thank the Syracuse Soft Matter Program for support and the KITP for its hospitality during the completion of some of the work.

\appendix
\begin{widetext}
\section{Steady state and linearization for $\lambda_2,\lambda_3\neq0$}
\label{sec:lambda}

Including the two additional $\lambda_2\b{p}\del\cdot\b{p}$ and $\lambda_3\del|\b{p}|^2$ nonlinearities, the equation for the polarization order parameter is modified to
\begin{equation}
		\partial_tp^{\mu}+\lambda p^{\nu}\del_{\nu}p^{\mu}=\left[a(\rho-\rho_c)-b\ g_{\alpha\beta}p^{\alpha}p^{\beta}\right]p^{\mu}+\lambda_2p^{\mu}\del_{\nu}p^{\nu}+\lambda_{3}\del^{\mu}(p^{\nu}p_{\nu})-v_1\del^{\mu}\rho.
\end{equation}
\end{widetext}
We don't include the viscous terms $\nu,\nu'$, consistent with our approximations in the main text. For the azimuthally symmetric ansatz (on the sphere and the catenoid), $\del_{\mu}p_{ss}^{\mu}=0$ identically and hence the $\lambda_2$ terms do not affect the steady state profile. On the contrary, the $\lambda_3\del|\b{p}|^2$ term acts as an additional polarization dependent contribution to the scalar pressure $P\sim v_1\rho-\lambda_3|\b{p}|^2$, which if large can lead to density and splay instabilities \cite{gopinath2012dynamical}. We shall disregard this and only work in the regime where $\lambda_3$ is not large enough to destabilize the entire system. Including it, the steady state equations on the sphere (Eqs. (\ref{eq:ptheta}), (\ref{eq:pphi})) get modified to 
\begin{gather}
	\lambda\sin\theta\cos\theta(p^{\phi}_{ss})^2=\dfrac{v_1}{R^2}\partial_{\theta}\rho_{ss}-\lambda_3\partial_{\theta}(\sin^2\theta(p^{\phi}_{ss})^2),\\
	p^{\phi}_{ss}\left[a(\rho_{ss}-\rho_c)-2bR^2\sin^2\theta(p^{\phi}_{ss})^2\right]=0.
\end{gather}
Once again, setting $X(\theta)=\rho_{ss}(\theta)-\rho_c$, we get the same equation as before, only now with a modified coefficient depending on $\lambda_3$.
\begin{equation}
	\dfrac{\dd X}{\dd\theta}=\dfrac{\lambda a}{v_1 b-a\lambda_3}\cot\theta X\implies X(\theta)=X(\pi/2)\sin^{\eta'}\theta,
\end{equation}
with the exponent now changed to $\eta'=\lambda a/(v_1b-a\lambda_3)$. For $\eta'>0$ (to have a physical density profile), we require $v_1b>a\lambda_3$, which is nothing but a condition to have stable pressure and a positive compressibility. Hence the effect of $\lambda_3\neq0$ is to only change the density profile through the $\eta$ exponent, the functional form remaining the same. This is true even for the catenoid where the exponent is the same as on the sphere and given by $\eta'=\lambda a/(v_1b-a\lambda_3)$. Note that this inhomogeneous profile does not exist for an equilbrium polar liquid crystal for which $\lambda=0$ (but possibly $\lambda_2,\lambda_3\neq0$ allowing for spontaneous splay \cite{kung2006hydrodynamics}), leading to $\eta'=0$. Hence the inhomogeneous steady state we obtain is only possible in an active system.

Linearizing about the steady state (on the sphere), now including the $\lambda_2$ and $\lambda_3$ terms, we get
\begin{align}
	&\partial_t\delta\rho+\partial_x\u+\partial_y\v=m\ \v,\\
	&\partial_t\u+(\bar{\lambda}-\bar{\lambda}_2-2\bar{\lambda}_3)\partial_x\u+\dfrac{\bar{v}_1}{\sin^2\theta_0}\partial_x\delta\rho=\nonumber\\
	&\quad\quad\alpha\delta\rho-\beta\u+\v\ m\left(\bar{\lambda}\dfrac{\eta'+2}{2}-\bar{\lambda}_2\right)+\bar{\lambda}_2\partial_y\v,\\
	&\partial_t\v+\bar{\lambda}\partial_x\u+\bar{v}_1\partial_y\delta\rho=-\left(2\bar{\lambda}+(4+\eta')\bar{\lambda}_3\right)\sin^2\theta_0 m\ \u\nonumber\\
	&\quad\quad\quad\quad\quad\quad\quad\quad\quad\quad+2\bar{\lambda}_3\sin^2\theta_0\partial_y\u.
\end{align}
We use the same notation as we used in the main text (along with $\bar{\lambda}_{2,3}=\lambda_{2,3} p_0$). At long times ($t\gg\beta^{-1}$), the polarization magnitude $\u$ is still a fast mode and to leading order, it gets slaved to the density fluctuations in the same fashion as before ($\u\simeq\alpha\delta\rho/\beta$). Consistent with the level of approximation used in the main text (Eq.~(\ref{eq:regime})), the long time and long wavelength dynamics is governed only by the two slow modes $\delta\rho$ and $\v$, with a dynamical matrix $D(\b{q})$ of the same form as found in the absence of $\lambda_2$ and $\lambda_3$.
\begin{equation}
	D(\b{q})=\left(\begin{matrix}
		s q_x & q_y+i m\\
			\bar{v}'_1 q_y-i\mu' m & 0
	\end{matrix}\right),
\end{equation}
with $s=\sgn(\alpha-\bar{\lambda}\beta)$ and the only modification being in the coefficients $\bar{v}'_1$ and $\mu'$.
\begin{gather}
	\bar{v}'_1=\bar{v}_1-2\bar{\lambda}_3\sin^2\theta_0\dfrac{\alpha}{\beta}=\dfrac{v_1b-a\lambda_3}{bR^2},\\
	\mu'=(2\bar{\lambda}+(4+\eta')\bar{\lambda}_3)\sin^2\theta_0\dfrac{\alpha}{\beta}=a\dfrac{\left[2\lambda+(4+\eta')\lambda_3\right]}{2bR^2}.
\end{gather}
It is easy to see now that the profile of the localized equatorial density mode on the sphere, which was $\propto\sin^{\mu'/\bar{v}'_1}\theta$ (for $\lambda_{2,3}=0$, $\mu/\bar{v}_1=\eta$ as given in Eq.~(\ref{eq:eqmode})) is no longer the same as the steady state density profile of the flock ($\sim\sin^{\eta'}\theta$), when $\lambda_3\neq0$.
\begin{equation}
	\dfrac{\mu'}{\bar{v}'_1}=\eta'\left[1+\left(2+\dfrac{\eta'}{2}\right)\dfrac{\lambda_3}{\lambda_1}\right]\neq\eta'.
\end{equation}
A similar result also holds true for the catenoid. Hence we find that, even upon including additional convective non-linearities (which are lower order in gradients compared to the viscous terms), all of the qualitative properties of the steady state and the topologically protected modes remains the same, with the only modification being a more detailed dependence of the localization length on some of the material parameters in the system.

\section{Polar flock on the catenoid}
\label{sec:catenoid}
For an azimuthally symmetric ordered steady state on the catenoid, just as on the sphere, we neglect the viscous and elastic stresses, and use the ansatz : $\rho=\rho_{ss}(y)$, $p^{y}=0$ and $p^{\phi}=p^{\phi}_{ss}(y)$. Plugging this into the Eqs.~(\ref{eq:cont}) and (\ref{eq:peq}), we find that the continuity equation is satisfied identically and Eq.~(\ref{eq:peq}) reduces to (for $\nu=0$)
\begin{gather}
	\lambda\tanh y(p^{\phi}_{ss})^2=\dfrac{v_1}{R^2\cosh^2y}\partial_y\rho_{ss},\\
	p^{\phi}_{ss}\left[a(\rho_{ss}-\rho_c)-bR^2\cosh^2y(p^{\phi}_{ss})^2\right]=0.
\end{gather}
Setting $X(y)=\rho_{ss}(y)-\rho_c$, we solve the equations in the same fashion as before to get $\partial_yX=\eta\tanh yX$, where $\eta=\lambda a/bv_1$ (the same exponent as on the sphere). The steady state density profile is then
\begin{equation}
	\rho_{ss}(y)=\rho_c+(\rho_{\mathrm{min}}-\rho_c)\cosh^{\eta} y.
\end{equation}
where $\rho_{\mathrm{min}}$ is the minimum density of the flock attained on the neck of the catenoid ($y=0$). Unlike the sphere, the catenoid is not a compact surface, so in reality one would have a finite sample with boundaries. The mean density $\rho_0$ is given by the spatial average of the steady state profile,
\begin{equation}
	\rho_0=\rho_c+(\rho_{\mathrm{min}}-\rho_c)\langle\cosh^{\eta}y\rangle,
\end{equation}
where $\langle\cdot\rangle$ denotes a spatial average over the entire surface. For a catenoid of height $L$ (euclidean height in the $z$-direction when embedded in $\mathbb{R}^3$) and radius of curvature $R$ at the minimal neck, we have
\begin{equation}
	\langle\cosh^{\eta}y\rangle\simeq\begin{cases}
		1+\dfrac{\eta}{6}\left(\dfrac{L}{R}\right)^2,\quad L/R\ll 1\\
		\dfrac{2^{1-\eta}}{2+\eta}e^{\eta L/R},\quad L/R\gg 1
	\end{cases}
\end{equation}
Writing $B_{\eta}=1/\langle\cosh^{\eta}y\rangle<1$, we obtain the density profile quoted in the main text (Eq.~(\ref{eq:catrho}))

We expect the viscous and elastic stresses to be less important on a weakly curved surface close to the neck, in particular when the characteristic scale of curvature ($\sim R$) is much greater than the equilibrium correlation length ($\xi\sim\sqrt{\nu/a(\rho_0-\rho_c)}$). Additionally the density and polarization (along with their gradients) grow larger as we go away from the neck. So close to the boundaries of a large sample, one would have to account for higher order nonlinearities along with the elastic stresses, which would then become important.

\subsection{Linearizing about the steady state}
One can perform the same kind of analysis as we did before for the flock on a sphere. Linearizing about the ordered flock, $\rho=\rho_{ss}(y)+\delta\rho$ and $\b{p}=\b{p}_{ss}(y)+\delta\b{p}$ within the tangent plane at a distance $y_0$ from the $y=0$ neck, we get (with $x=\phi$, $\u=\delta p^{\phi}$ and $\v=\delta p^{y}$ just as before)
\begin{align}
	&\partial_t\delta\rho+\partial_x\u+\partial_y\v+2\v\tanh y_0=0,\\
	&\partial_t\u+\lambda p_0\partial_x\u+\dfrac{v_1}{R^2\cosh^2y_0}\partial_x\delta\rho\nonumber\\
							   &\quad=p_0(a\delta\rho-2bR^2p_0\cosh^2y_0\ \u)-\v\dfrac{\lambda p_0(\eta+2)}{2}\tanh y_0,\\
			 &\partial_t\v+\lambda p_0\partial_x\v+\dfrac{v_1}{R^2\cosh^2y_0}\partial_y\delta\rho=2\u\lambda p_0\tanh y_0,
\end{align}
where $p_0=p^{\phi}_{ss}(y_0)$. Galilean boosting to a moving frame $x\to x-\lambda p_0 t$ and relabelling our parameters as before : $\bar{\lambda}=\lambda p_0$, $\alpha=a p_0>0$, $\beta=2bR^2p_0^2\cosh^2y_0>0$, $\bar{v}_1=v_1/(R^2\cosh^2y_0)$ and $m=-2\tanh y_0$. Having done this, all the arguments used in the case of the sphere apply here as well.

At long time ($t\gg\beta^{-1}$), the fast polarization magnitude $\u$ decays and is slaved to the density field $\u\simeq\alpha\delta\rho/\beta$ (to lowest order) and the slow dynamics at long wavelengths is dominated by
\begin{gather}
	\partial_t\delta\rho+\left(\alpha/\beta-\bar{\lambda}\right)\partial_x\delta\rho+\partial_y\v=m\v,\\
	\partial_t\v+\bar{v}_1\partial_y\delta\rho=-\mu m\delta\rho,
\end{gather}
only now with $\mu=\alpha\bar{\lambda}/\beta>0$. Hence at the same level of approximation used earlier for the sphere (neglecting viscous stresses and the parameter regime given in Eq.~(\ref{eq:regime})), the long time dynamics of sound excitations in a polar flock on a curved surface is generically described by equations of the form given above, or consequently by the linear dynamical matrix $D(\b{q})$ (Eq.~(\ref{eq:D})), possibly upto some coordinate rescaling.

As the only modifications are in the definitions of the parameters, many of the predictions made in the case of the sphere apply here too. In particular the sound mode spectrum is still gapped at $\b{q}=0$ for non-zero $m$ and the bands have a non-trivial topology given by the Chern numbers $C_{\pm}$ (see Eq.~(\ref{eq:chern})). As $m=0$ at the neck of the catenoid ($y=0$), changing sign on either side, we have one topologically protected mode localized at the neck. A consequence of the negative curvature of the surface is that, in contrast to the sphere, $m<0$ for $y>0$. Due to this, the edge mode takes on a different form (in the comoving frame)
\begin{equation}
|\delta\Psi_{\mathrm{edge}}\rangle=\psi_0 e^{\int^y_0m(y')\dd y'+iq_x x}\left(\begin{matrix} 0\\ 1\end{matrix}\right).
\end{equation}
Now the edge mode is a localized mode of transverse goldstone fluctuations with density fluctuations completely absent. Additionally, the edge mode spectrum is $\omega_{\mathrm{edge}}=0$ to lowest order in $q_x$ implying that the edge mode is stationary in the comoving (with speed $\sim\lambda p_0$) frame. This too connects the two bulk bands and is topologically protected. Using $m(y)=-2\tanh y$ for the catenoid, this gives the profile of the localized mode, the lab frame version of which is quoted in the main text (Eq.~(\ref{eq:vedge})).
%\bibliography{refs}

\begin{thebibliography}{84}
\providecommand{\natexlab}[1]{#1}
\providecommand{\url}[1]{\texttt{#1}}
\expandafter\ifx\csname urlstyle\endcsname\relax
  \providecommand{\doi}[1]{doi: #1}\else
  \providecommand{\doi}{doi: \begingroup \urlstyle{rm}\Url}\fi

\bibitem[Vicsek et~al.(1995)Vicsek, Czir{\'o}k, Ben-Jacob, Cohen, and
  Shochet]{vicsek1995novel}
Tam{\'a}s Vicsek, Andr{\'a}s Czir{\'o}k, Eshel Ben-Jacob, Inon Cohen, and Ofer
  Shochet.
\newblock Novel type of phase transition in a system of self-driven particles.
\newblock \emph{Physical review letters}, 75\penalty0 (6):\penalty0 1226, 1995.

\bibitem[Toner et~al.(2005)Toner, Tu, and Ramaswamy]{toner2005hydrodynamics}
John Toner, Yuhai Tu, and Sriram Ramaswamy.
\newblock Hydrodynamics and phases of flocks.
\newblock \emph{Annals of Physics}, 318\penalty0 (1):\penalty0 170--244, 2005.

\bibitem[Ramaswamy(2010)]{ramaswamy2010mechanics}
Sriram Ramaswamy.
\newblock The mechanics and statistics of active matter.
\newblock \emph{Annu. Rev. Condens. Matter Phys.}, 1\penalty0 (1):\penalty0
  323--345, 2010.

\bibitem[Marchetti et~al.(2013)Marchetti, Joanny, Ramaswamy, Liverpool, Prost,
  Rao, and Simha]{marchetti2013hydrodynamics}
M~Cristina Marchetti, JF~Joanny, S~Ramaswamy, TB~Liverpool, J~Prost, Madan Rao,
  and R~Aditi Simha.
\newblock Hydrodynamics of soft active matter.
\newblock \emph{Reviews of Modern Physics}, 85\penalty0 (3):\penalty0 1143,
  2013.

\bibitem[Cavagna et~al.(2010)Cavagna, Cimarelli, Giardina, Parisi, Santagati,
  Stefanini, and Viale]{cavagna2010scale}
Andrea Cavagna, Alessio Cimarelli, Irene Giardina, Giorgio Parisi, Raffaele
  Santagati, Fabio Stefanini, and Massimiliano Viale.
\newblock Scale-free correlations in starling flocks.
\newblock \emph{Proceedings of the National Academy of Sciences}, 107\penalty0
  (26):\penalty0 11865--11870, 2010.

\bibitem[Sokolov et~al.(2007)Sokolov, Aranson, Kessler, and
  Goldstein]{sokolov2007concentration}
Andrey Sokolov, Igor~S Aranson, John~O Kessler, and Raymond~E Goldstein.
\newblock Concentration dependence of the collective dynamics of swimming
  bacteria.
\newblock \emph{Physical Review Letters}, 98\penalty0 (15):\penalty0 158102,
  2007.

\bibitem[Szabo et~al.(2006)Szabo, Sz{\"o}ll{\"o}si, G{\"o}nci, Jur{\'a}nyi,
  Selmeczi, and Vicsek]{szabo2006phase}
Balint Szabo, GJ~Sz{\"o}ll{\"o}si, B~G{\"o}nci, Zs~Jur{\'a}nyi, David Selmeczi,
  and Tam{\'a}s Vicsek.
\newblock Phase transition in the collective migration of tissue cells:
  experiment and model.
\newblock \emph{Physical Review E}, 74\penalty0 (6):\penalty0 061908, 2006.

\bibitem[Schaller et~al.(2010)Schaller, Weber, Semmrich, Frey, and
  Bausch]{schaller2010polar}
Volker Schaller, Christoph Weber, Christine Semmrich, Erwin Frey, and Andreas~R
  Bausch.
\newblock Polar patterns of driven filaments.
\newblock \emph{Nature}, 467\penalty0 (7311):\penalty0 73--77, 2010.

\bibitem[Sanchez et~al.(2012)Sanchez, Chen, DeCamp, Heymann, and
  Dogic]{sanchez2012spontaneous}
Tim Sanchez, Daniel~TN Chen, Stephen~J DeCamp, Michael Heymann, and Zvonimir
  Dogic.
\newblock Spontaneous motion in hierarchically assembled active matter.
\newblock \emph{Nature}, 491\penalty0 (7424):\penalty0 431--434, 2012.

\bibitem[Keber et~al.(2014)Keber, Loiseau, Sanchez, DeCamp, Giomi, Bowick,
  Marchetti, Dogic, and Bausch]{keber2014topology}
Felix~C Keber, Etienne Loiseau, Tim Sanchez, Stephen~J DeCamp, Luca Giomi,
  Mark~J Bowick, M~Cristina Marchetti, Zvonimir Dogic, and Andreas~R Bausch.
\newblock Topology and dynamics of active nematic vesicles.
\newblock \emph{Science}, 345\penalty0 (6201):\penalty0 1135--1139, 2014.

\bibitem[Deseigne et~al.(2010)Deseigne, Dauchot, and
  Chat{\'e}]{deseigne2010collective}
Julien Deseigne, Olivier Dauchot, and Hugues Chat{\'e}.
\newblock Collective motion of vibrated polar disks.
\newblock \emph{Physical review letters}, 105\penalty0 (9):\penalty0 098001,
  2010.

\bibitem[Narayan et~al.(2007)Narayan, Ramaswamy, and Menon]{narayan2007long}
Vijay Narayan, Sriram Ramaswamy, and Narayanan Menon.
\newblock Long-lived giant number fluctuations in a swarming granular nematic.
\newblock \emph{Science}, 317\penalty0 (5834):\penalty0 105--108, 2007.

\bibitem[Palacci et~al.(2013)Palacci, Sacanna, Steinberg, Pine, and
  Chaikin]{palacci2013living}
Jeremie Palacci, Stefano Sacanna, Asher~Preska Steinberg, David~J Pine, and
  Paul~M Chaikin.
\newblock Living crystals of light-activated colloidal surfers.
\newblock \emph{Science}, 339\penalty0 (6122):\penalty0 936--940, 2013.

\bibitem[Streichan et~al.(2017)Streichan, Lefebvre, Noll, Wieschaus, and
  Shraiman]{streichan2017quantification}
Sebastian~J Streichan, Matthew~F Lefebvre, Nicholas Noll, Eric~F Wieschaus, and
  Boris~I Shraiman.
\newblock Quantification of myosin distribution predicts global morphogenetic
  flow in the fly embryo.
\newblock \emph{arXiv preprint arXiv:1701.07100}, 2017.

\bibitem[Vasiev et~al.(2010)Vasiev, Balter, Chaplain, Glazier, and
  Weijer]{vasiev2010modeling}
Bakhtier Vasiev, Ariel Balter, Mark Chaplain, James~A Glazier, and Cornelis~J
  Weijer.
\newblock Modeling gastrulation in the chick embryo: formation of the primitive
  streak.
\newblock \emph{PLoS One}, 5\penalty0 (5):\penalty0 e10571, 2010.

\bibitem[Ewald et~al.(2008)Ewald, Brenot, Duong, Chan, and
  Werb]{ewald2008collective}
Andrew~J Ewald, Audrey Brenot, Myhanh Duong, Bianca~S Chan, and Zena Werb.
\newblock Collective epithelial migration and cell rearrangements drive mammary
  branching morphogenesis.
\newblock \emph{Developmental cell}, 14\penalty0 (4):\penalty0 570--581, 2008.

\bibitem[Fatehullah et~al.(2013)Fatehullah, Appleton, and
  N{\"a}thke]{fatehullah2013cell}
Aliya Fatehullah, Paul~L Appleton, and Inke~S N{\"a}thke.
\newblock Cell and tissue polarity in the intestinal tract during
  tumourigenesis: cells still know the right way up, but tissue organization is
  lost.
\newblock \emph{Phil. Trans. R. Soc. B}, 368\penalty0 (1629):\penalty0
  20130014, 2013.

\bibitem[Ritsma et~al.(2014)Ritsma, Ellenbroek, Zomer, Snippert, de~Sauvage,
  Simons, Clevers, and van Rheenen]{ritsma2014intestinal}
Laila Ritsma, Saskia~IJ Ellenbroek, Anoek Zomer, Hugo~J Snippert, Frederic~J
  de~Sauvage, Benjamin~D Simons, Hans Clevers, and Jacco van Rheenen.
\newblock Intestinal crypt homeostasis revealed at single-stem-cell level by in
  vivo live imaging.
\newblock \emph{Nature}, 507\penalty0 (7492):\penalty0 362--365, 2014.

\bibitem[Collinson et~al.(2002)Collinson, Morris, Reid, Ramaesh, Keighren,
  Flockhart, Hill, Tan, Ramaesh, Dhillon, and West]{collinson2002clonal}
J.~Martin Collinson, Lucy Morris, Alasdair~I. Reid, Thaya Ramaesh, Margaret~A.
  Keighren, Jean~H. Flockhart, Robert~E. Hill, Seong-Seng Tan, Kanna Ramaesh,
  Baljean Dhillon, and John~D. West.
\newblock Clonal analysis of patterns of growth, stem cell activity, and cell
  movement during the development and maintenance of the murine corneal
  epithelium.
\newblock \emph{Developmental Dynamics}, 224\penalty0 (4):\penalty0 432--440,
  2002.
\newblock ISSN 1097-0177.
\newblock \doi{10.1002/dvdy.10124}.

\bibitem[Mishra et~al.(2012)Mishra, Huang, Srivastava, Srinivasan, Sevugan,
  Shlomovitz, Gov, Rao, and Balasubramanian]{mishra2012cylindrical}
Mithilesh Mishra, Yinyi Huang, Pragya Srivastava, Ramanujam Srinivasan,
  Mayalagu Sevugan, Roie Shlomovitz, Nir Gov, Madan Rao, and Mohan
  Balasubramanian.
\newblock Cylindrical cellular geometry ensures fidelity of division site
  placement in fission yeast.
\newblock \emph{J Cell Sci}, 125\penalty0 (16):\penalty0 3850--3857, 2012.

\bibitem[Yevick et~al.(2015)Yevick, Duclos, Bonnet, and
  Silberzan]{yevick2015architecture}
Hannah~G Yevick, Guillaume Duclos, Isabelle Bonnet, and Pascal Silberzan.
\newblock Architecture and migration of an epithelium on a cylindrical wire.
\newblock \emph{Proceedings of the National Academy of Sciences}, 112\penalty0
  (19):\penalty0 5944--5949, 2015.

\bibitem[Fily et~al.(2014)Fily, Baskaran, and Hagan]{fily2014dynamics}
Yaouen Fily, Aparna Baskaran, and Michael~F Hagan.
\newblock Dynamics of self-propelled particles under strong confinement.
\newblock \emph{Soft matter}, 10\penalty0 (30):\penalty0 5609--5617, 2014.

\bibitem[Fily et~al.(2015)Fily, Baskaran, and Hagan]{fily2015dynamics}
Yaouen Fily, Aparna Baskaran, and Michael~F Hagan.
\newblock Dynamics and density distribution of strongly confined noninteracting
  nonaligning self-propelled particles in a nonconvex boundary.
\newblock \emph{Physical Review E}, 91\penalty0 (1):\penalty0 012125, 2015.

\bibitem[Fily et~al.(2016{\natexlab{a}})Fily, Baskaran, and
  Hagan]{fily2016equilibrium}
Yaouen Fily, Aparna Baskaran, and Michael~F Hagan.
\newblock Equilibrium mappings in polar-isotropic confined active particles.
\newblock \emph{arXiv preprint arXiv:1612.08719}, 2016{\natexlab{a}}.

\bibitem[Nikola et~al.(2016)Nikola, Solon, Kafri, Kardar, Tailleur, and
  Voituriez]{nikola2016active}
Nikolai Nikola, Alexandre~P Solon, Yariv Kafri, Mehran Kardar, Julien Tailleur,
  and Rapha{\"e}l Voituriez.
\newblock Active particles with soft and curved walls: Equation of state,
  ratchets, and instabilities.
\newblock \emph{Physical Review Letters}, 117\penalty0 (9):\penalty0 098001,
  2016.

\bibitem[Smallenburg and L{\"o}wen(2015)]{smallenburg2015swim}
Frank Smallenburg and Hartmut L{\"o}wen.
\newblock Swim pressure on walls with curves and corners.
\newblock \emph{Physical Review E}, 92\penalty0 (3):\penalty0 032304, 2015.

\bibitem[Green et~al.(2016)Green, Toner, and Vitelli]{green2016geometry}
Richard Green, John Toner, and Vincenzo Vitelli.
\newblock The geometry of threshold-less active flow in nematic microfluidics.
\newblock \emph{arXiv preprint arXiv:1602.00561}, 2016.

\bibitem[Fily et~al.(2016{\natexlab{b}})Fily, Baskaran, and
  Hagan]{fily2016active}
Yaouen Fily, Aparna Baskaran, and Michael~F Hagan.
\newblock Active particles on curved surfaces.
\newblock \emph{arXiv preprint arXiv:1601.00324}, 2016{\natexlab{b}}.

\bibitem[Sknepnek and Henkes(2015)]{sknepnek2015active}
Rastko Sknepnek and Silke Henkes.
\newblock Active swarms on a sphere.
\newblock \emph{Physical Review E}, 91\penalty0 (2):\penalty0 022306, 2015.

\bibitem[Nelson(2002)]{nelson2002defects}
David~R Nelson.
\newblock \emph{Defects and geometry in condensed matter physics}.
\newblock Cambridge University Press, 2002.

\bibitem[Maitra et~al.(2014)Maitra, Srivastava, Rao, and
  Ramaswamy]{maitra2014activating}
Ananyo Maitra, Pragya Srivastava, Madan Rao, and Sriram Ramaswamy.
\newblock Activating membranes.
\newblock \emph{Physical review letters}, 112\penalty0 (25):\penalty0 258101,
  2014.

\bibitem[Mallory et~al.(2014)Mallory, Valeriani, and
  Cacciuto]{mallory2014curvature}
SA~Mallory, C~Valeriani, and A~Cacciuto.
\newblock Curvature-induced activation of a passive tracer in an active bath.
\newblock \emph{Physical Review E}, 90\penalty0 (3):\penalty0 032309, 2014.

\bibitem[Souslov et~al.(2016)Souslov, van Zuiden, Bartolo, and
  Vitelli]{souslov2016topological}
Anton Souslov, Benjamin~C van Zuiden, Denis Bartolo, and Vincenzo Vitelli.
\newblock Topological sound in active-liquid metamaterials.
\newblock \emph{arXiv preprint arXiv:1610.06873}, 2016.

\bibitem[Yang et~al.(2015)Yang, Gao, Shi, Lin, Gao, Chong, and
  Zhang]{yang2015topological}
Zhaoju Yang, Fei Gao, Xihang Shi, Xiao Lin, Zhen Gao, Yidong Chong, and Baile
  Zhang.
\newblock Topological acoustics.
\newblock \emph{Physical review letters}, 114\penalty0 (11):\penalty0 114301,
  2015.

\bibitem[Prodan and Prodan(2009)]{prodan2009topological}
Emil Prodan and Camelia Prodan.
\newblock Topological phonon modes and their role in dynamic instability of
  microtubules.
\newblock \emph{Physical review letters}, 103\penalty0 (24):\penalty0 248101,
  2009.

\bibitem[Kane and Lubensky(2014)]{kane2014topological}
CL~Kane and TC~Lubensky.
\newblock Topological boundary modes in isostatic lattices.
\newblock \emph{Nature Physics}, 10\penalty0 (1):\penalty0 39--45, 2014.

\bibitem[Raghu and Haldane(2008)]{raghu2008analogs}
S~Raghu and FDM Haldane.
\newblock Analogs of quantum-hall-effect edge states in photonic crystals.
\newblock \emph{Physical Review A}, 78\penalty0 (3):\penalty0 033834, 2008.

\bibitem[Hasan and Kane(2010)]{hasan2010colloquium}
M~Zahid Hasan and Charles~L Kane.
\newblock Colloquium: topological insulators.
\newblock \emph{Reviews of Modern Physics}, 82\penalty0 (4):\penalty0 3045,
  2010.

\bibitem[Delplace et~al.(2017)Delplace, Marston, and Venaille]{marston2017}
Pierre Delplace, JB~Marston, and Antoine Venaille.
\newblock Topological origin of geophysical waves.
\newblock \emph{arXiv preprint arXiv:1702.07583}, 2017.

\bibitem[Srivastava et~al.(2013)Srivastava, Shlomovitz, Gov, and
  Rao]{srivastava2013patterning}
Pragya Srivastava, Roie Shlomovitz, Nir~S Gov, and Madan Rao.
\newblock Patterning of polar active filaments on a tense cylindrical membrane.
\newblock \emph{Physical review letters}, 110\penalty0 (16):\penalty0 168104,
  2013.

\bibitem[Eisenberg and Guy(1979)]{eisenberg1979proof}
Murray Eisenberg and Robert Guy.
\newblock A proof of the hairy ball theorem.
\newblock \emph{The American Mathematical Monthly}, 86\penalty0 (7):\penalty0
  571--574, 1979.

\bibitem[Tu et~al.(1998)Tu, Toner, and Ulm]{tu1998sound}
Yuhai Tu, John Toner, and Markus Ulm.
\newblock Sound waves and the absence of galilean invariance in flocks.
\newblock \emph{Physical review letters}, 80\penalty0 (21):\penalty0 4819,
  1998.

\bibitem[Toner and Tu(1995)]{toner1995long}
John Toner and Yuhai Tu.
\newblock Long-range order in a two-dimensional dynamical xy model: how birds
  fly together.
\newblock \emph{Physical Review Letters}, 75\penalty0 (23):\penalty0 4326,
  1995.

\bibitem[Toner and Tu(1998)]{toner1998flocks}
John Toner and Yuhai Tu.
\newblock Flocks, herds, and schools: A quantitative theory of flocking.
\newblock \emph{Physical review E}, 58\penalty0 (4):\penalty0 4828, 1998.

\bibitem[do~Carmo~Valero(1992)]{do1992riemannian}
Manfredo~Perdigao do~Carmo~Valero.
\newblock \emph{Riemannian geometry}.
\newblock Prentice-Hall, 1992.

\bibitem[Yang et~al.(2014)Yang, Manning, and Marchetti]{yang2014aggregation}
Xingbo Yang, M~Lisa Manning, and M~Cristina Marchetti.
\newblock Aggregation and segregation of confined active particles.
\newblock \emph{Soft matter}, 10\penalty0 (34):\penalty0 6477--6484, 2014.

\bibitem[Takatori et~al.(2014)Takatori, Yan, and Brady]{takatori2014swim}
Sho~C Takatori, Wen Yan, and John~F Brady.
\newblock Swim pressure: stress generation in active matter.
\newblock \emph{Physical review letters}, 113\penalty0 (2):\penalty0 028103,
  2014.

\bibitem[Toner(2012)]{toner2012reanalysis}
John Toner.
\newblock Reanalysis of the hydrodynamic theory of fluid, polar-ordered flocks.
\newblock \emph{Physical Review E}, 86\penalty0 (3):\penalty0 031918, 2012.

\bibitem[Gopinath et~al.(2012)Gopinath, Hagan, Marchetti, and
  Baskaran]{gopinath2012dynamical}
Arvind Gopinath, Michael~F Hagan, M~Cristina Marchetti, and Aparna Baskaran.
\newblock Dynamical self-regulation in self-propelled particle flows.
\newblock \emph{Physical Review E}, 85\penalty0 (6):\penalty0 061903, 2012.

\bibitem[Bertin et~al.(2009)Bertin, Droz, and
  Gr{\'e}goire]{bertin2009hydrodynamic}
Eric Bertin, Michel Droz, and Guillaume Gr{\'e}goire.
\newblock Hydrodynamic equations for self-propelled particles: microscopic
  derivation and stability analysis.
\newblock \emph{Journal of Physics A: Mathematical and Theoretical},
  42\penalty0 (44):\penalty0 445001, 2009.

\bibitem[Solon et~al.(2015)Solon, Caussin, Bartolo, Chat{\'e}, and
  Tailleur]{solon2015pattern}
Alexandre~P Solon, Jean-Baptiste Caussin, Denis Bartolo, Hugues Chat{\'e}, and
  Julien Tailleur.
\newblock Pattern formation in flocking models: A hydrodynamic description.
\newblock \emph{Physical Review E}, 92\penalty0 (6):\penalty0 062111, 2015.

\bibitem[Caussin et~al.(2014)Caussin, Solon, Peshkov, Chat{\'e}, Dauxois,
  Tailleur, Vitelli, and Bartolo]{caussin2014emergent}
Jean-Baptiste Caussin, Alexandre Solon, Anton Peshkov, Hugues Chat{\'e},
  Thierry Dauxois, Julien Tailleur, Vincenzo Vitelli, and Denis Bartolo.
\newblock Emergent spatial structures in flocking models: a dynamical system
  insight.
\newblock \emph{Physical review letters}, 112\penalty0 (14):\penalty0 148102,
  2014.

\bibitem[Mishra et~al.(2010)Mishra, Baskaran, and
  Marchetti]{mishra2010fluctuations}
Shradha Mishra, Aparna Baskaran, and M~Cristina Marchetti.
\newblock Fluctuations and pattern formation in self-propelled particles.
\newblock \emph{Physical Review E}, 81\penalty0 (6):\penalty0 061916, 2010.

\bibitem[Bertin et~al.(2006)Bertin, Droz, and
  Gr{\'e}goire]{bertin2006boltzmann}
Eric Bertin, Michel Droz, and Guillaume Gr{\'e}goire.
\newblock Boltzmann and hydrodynamic description for self-propelled particles.
\newblock \emph{Physical Review E}, 74\penalty0 (2):\penalty0 022101, 2006.

\bibitem[De~Gennes and Prost(1993)]{de1997physics}
Pierre~Gilles De~Gennes and Jacques Prost.
\newblock \emph{The physics of liquid crystals}.
\newblock Clarendon Press, Oxford, 1993.

\bibitem[Kung et~al.(2006)Kung, Marchetti, and Saunders]{kung2006hydrodynamics}
William Kung, M~Cristina Marchetti, and Karl Saunders.
\newblock Hydrodynamics of polar liquid crystals.
\newblock \emph{Physical Review E}, 73\penalty0 (3):\penalty0 031708, 2006.

\bibitem[Baskaran and Marchetti(2008)]{baskaran2008hydrodynamics}
Aparna Baskaran and M~Cristina Marchetti.
\newblock Hydrodynamics of self-propelled hard rods.
\newblock \emph{Physical Review E}, 77\penalty0 (1):\penalty0 011920, 2008.

\bibitem[Martin et~al.(1972)Martin, Parodi, and Pershan]{martin1972unified}
PC~Martin, O~Parodi, and Peter~S Pershan.
\newblock Unified hydrodynamic theory for crystals, liquid crystals, and normal
  fluids.
\newblock \emph{Physical Review A}, 6\penalty0 (6):\penalty0 2401, 1972.

\bibitem[Gill(2016)]{gill2016atmosphere}
Adrian~E Gill.
\newblock \emph{Atmosphere - ocean dynamics}.
\newblock Elsevier, 2016.

\bibitem[Berry(1984)]{berry1984quantal}
Michael~V Berry.
\newblock Quantal phase factors accompanying adiabatic changes.
\newblock In \emph{Proceedings of the Royal Society of London A: Mathematical,
  Physical and Engineering Sciences}, volume 392, pages 45--57. The Royal
  Society, 1984.

\bibitem[Nakahara(2003)]{nakahara2003geometry}
Mikio Nakahara.
\newblock \emph{Geometry, topology and physics}.
\newblock Taylor \& Francis, 2003.

\bibitem[Haldane(1988)]{haldane1988model}
F~Duncan~M Haldane.
\newblock Model for a quantum hall effect without landau levels:
  Condensed-matter realization of the" parity anomaly".
\newblock \emph{Physical Review Letters}, 61\penalty0 (18):\penalty0 2015,
  1988.

\bibitem[Ryu et~al.(2010)Ryu, Schnyder, Furusaki, and
  Ludwig]{ryu2010topological}
Shinsei Ryu, Andreas~P Schnyder, Akira Furusaki, and Andreas~WW Ludwig.
\newblock Topological insulators and superconductors: tenfold way and
  dimensional hierarchy.
\newblock \emph{New Journal of Physics}, 12\penalty0 (6):\penalty0 065010,
  2010.

\bibitem[Leykam et~al.(2017)Leykam, Bliokh, Huang, Chong, and
  Nori]{leykam2017edge}
Daniel Leykam, Konstantin~Y Bliokh, Chunli Huang, YD~Chong, and Franco Nori.
\newblock Edge modes, degeneracies, and topological numbers in non-hermitian
  systems.
\newblock \emph{Physical Review Letters}, 118\penalty0 (4):\penalty0 040401,
  2017.

\bibitem[Jackiw and Rebbi(1976)]{jackiw1976solitons}
Roman Jackiw and Claudio Rebbi.
\newblock Solitons with fermion number $1/2$.
\newblock \emph{Physical Review D}, 13\penalty0 (12):\penalty0 3398, 1976.

\bibitem[Bowick and Giomi(2009)]{bowick2009two}
Mark~J Bowick and Luca Giomi.
\newblock Two-dimensional matter: order, curvature and defects.
\newblock \emph{Advances in Physics}, 58\penalty0 (5):\penalty0 449--563, 2009.

\bibitem[Turner et~al.(2010)Turner, Vitelli, and Nelson]{turner2010vortices}
Ari~M Turner, Vincenzo Vitelli, and David~R Nelson.
\newblock Vortices on curved surfaces.
\newblock \emph{Reviews of Modern Physics}, 82\penalty0 (2):\penalty0 1301,
  2010.

\bibitem[Bowick et~al.(2000)Bowick, Nelson, and
  Travesset]{bowick2000interacting}
Mark~J Bowick, David~R Nelson, and Alex Travesset.
\newblock Interacting topological defects on frozen topographies.
\newblock \emph{Physical Review B}, 62\penalty0 (13):\penalty0 8738, 2000.

\bibitem[Bausch et~al.(2003)Bausch, Bowick, Cacciuto, Dinsmore, Hsu, Nelson,
  Nikolaides, Travesset, and Weitz]{bausch2003grain}
AR~Bausch, MJ~Bowick, A~Cacciuto, AD~Dinsmore, MF~Hsu, DR~Nelson,
  MG~Nikolaides, A~Travesset, and DA~Weitz.
\newblock Grain boundary scars and spherical crystallography.
\newblock \emph{Science}, 299\penalty0 (5613):\penalty0 1716--1718, 2003.

\bibitem[Irvine et~al.(2010)Irvine, Vitelli, and Chaikin]{irvine2010pleats}
William~TM Irvine, Vincenzo Vitelli, and Paul~M Chaikin.
\newblock Pleats in crystals on curved surfaces.
\newblock \emph{Nature}, 468\penalty0 (7326):\penalty0 947--951, 2010.

\bibitem[Bowick and Yao(2011)]{bowick2011crystalline}
Mark~J Bowick and Zhenwei Yao.
\newblock Crystalline order on catenoidal capillary bridges.
\newblock \emph{EPL (Europhysics Letters)}, 93\penalty0 (3):\penalty0 36001,
  2011.

\bibitem[Bendito et~al.(2013)Bendito, Bowick, Medina, and
  Yao]{bendito2013crystalline}
Enrique Bendito, Mark~J Bowick, Agustin Medina, and Zhenwei Yao.
\newblock Crystalline particle packings on constant mean curvature (delaunay)
  surfaces.
\newblock \emph{Physical Review E}, 88\penalty0 (1):\penalty0 012405, 2013.

\bibitem[Nelson and Peliti(1987)]{nelson1987fluctuations}
DR~Nelson and L~Peliti.
\newblock Fluctuations in membranes with crystalline and hexatic order.
\newblock \emph{Journal de Physique}, 48\penalty0 (7):\penalty0 1085--1092,
  1987.

\bibitem[Seung and Nelson(1988)]{seung1988defects}
HS~Seung and David~R Nelson.
\newblock Defects in flexible membranes with crystalline order.
\newblock \emph{Physical Review A}, 38\penalty0 (2):\penalty0 1005, 1988.

\bibitem[Xing et~al.(2012)Xing, Shin, Bowick, Yao, Jia, and
  Li]{xing2012morphology}
Xiangjun Xing, Homin Shin, Mark~J Bowick, Zhenwei Yao, Lin Jia, and Min-Hui Li.
\newblock Morphology of nematic and smectic vesicles.
\newblock \emph{Proceedings of the National Academy of Sciences}, 109\penalty0
  (14):\penalty0 5202--5206, 2012.

\bibitem[Ramakrishnan et~al.(2012)Ramakrishnan, Ipsen, and
  Kumar]{ramakrishnan2012role}
N~Ramakrishnan, John~H Ipsen, and PB~Sunil Kumar.
\newblock Role of disclinations in determining the morphology of deformable
  fluid interfaces.
\newblock \emph{Soft Matter}, 8\penalty0 (11):\penalty0 3058--3061, 2012.

\bibitem[Mesarec et~al.(2017)Mesarec, Kurioz, Igli{$\check{\mathrm{c}}$},
  G{\'o}{\'z}d{\'z}, and Kralj]{mesarec2017curvature}
Luka Mesarec, Pavlo Kurioz, Ale{$\check{\mathrm{s}}$}
  Igli{$\check{\mathrm{c}}$}, Wojciech G{\'o}{\'z}d{\'z}, and Samo Kralj.
\newblock Curvature-controlled topological defects.
\newblock \emph{Crystals}, 7\penalty0 (6):\penalty0 153, 2017.

\bibitem[Nelson(1983)]{nelson1983liquids}
David~R Nelson.
\newblock Liquids and glasses in spaces of incommensurate curvature.
\newblock \emph{Physical Review Letters}, 50\penalty0 (13):\penalty0 982, 1983.

\bibitem[Modes and Kamien(2007)]{modes2007hard}
Carl~D Modes and Randall~D Kamien.
\newblock Hard disks on the hyperbolic plane.
\newblock \emph{Physical review letters}, 99\penalty0 (23):\penalty0 235701,
  2007.

\bibitem[Vitelli and Turner(2004)]{vitelli2004anomalous}
Vincenzo Vitelli and Ari~M Turner.
\newblock Anomalous coupling between topological defects and curvature.
\newblock \emph{Physical review letters}, 93\penalty0 (21):\penalty0 215301,
  2004.

\bibitem[Nash et~al.(2015)Nash, Kleckner, Read, Vitelli, Turner, and
  Irvine]{nash2015topological}
Lisa~M Nash, Dustin Kleckner, Alismari Read, Vincenzo Vitelli, Ari~M Turner,
  and William~TM Irvine.
\newblock Topological mechanics of gyroscopic metamaterials.
\newblock \emph{Proceedings of the National Academy of Sciences}, 112\penalty0
  (47):\penalty0 14495--14500, 2015.

\bibitem[Serra-Picamal et~al.(2012)Serra-Picamal, Conte, Vincent, Anon, Tambe,
  Bazellieres, Butler, Fredberg, and Trepat]{serra2012mechanical}
Xavier Serra-Picamal, Vito Conte, Romaric Vincent, Ester Anon, Dhananjay~T
  Tambe, Elsa Bazellieres, James~P Butler, Jeffrey~J Fredberg, and Xavier
  Trepat.
\newblock Mechanical waves during tissue expansion.
\newblock \emph{Nature Physics}, 8\penalty0 (8):\penalty0 628--634, 2012.

\bibitem[Holmberg et~al.(2017)Holmberg, Seidelin, Yin, Mead, Tong, Li, Karp,
  and Nielsen]{Holmberge201607260}
Fredrik~EO Holmberg, Jakob~B Seidelin, Xiaolei Yin, Benjamin~E Mead, Zhixiang
  Tong, Yuan Li, Jeffrey~M Karp, and Ole~H Nielsen.
\newblock Culturing human intestinal stem cells for regenerative applications
  in the treatment of inflammatory bowel disease.
\newblock \emph{EMBO Molecular Medicine}, 2017.
\newblock ISSN 1757-4676.
\newblock \doi{10.15252/emmm.201607260}.

\bibitem[Hopyan et~al.(2011)Hopyan, Sharpe, and Yang]{hopyan2011budding}
Sevan Hopyan, James Sharpe, and Yingzi Yang.
\newblock Budding behaviors: Growth of the limb as a model of morphogenesis.
\newblock \emph{Developmental Dynamics}, 240\penalty0 (5):\penalty0 1054--1062,
  2011.

\end{thebibliography}
%\bibliographystyle{unsrtnat}

\end{document}